\documentclass[fleqn,usenatbib]{mnras}

\usepackage{newtxtext,newtxmath}

\usepackage[T1]{fontenc}
\usepackage{ae,aecompl}

\usepackage{graphicx}	
\usepackage{amsmath}	
\usepackage{amssymb}	

\title[2.5-D phase curve retrievals of WASP-43b]{2.5-D retrieval of atmospheric properties from exoplanet phase curves: Application to WASP-43b observations}

\author[Irwin et al.]{
Patrick G. J. Irwin,$^{1}$\thanks{E-mail: patrick.irwin@physics.ox.ac.uk}
Vivien Parmentier,$^{1}$
Jake Taylor,$^{1}$
Jo Barstow,$^{2}$
\newauthor
Suzanne Aigrain,$^{1}$
Elspeth K. H. Lee,$^{1}$
and Ryan Garland,$^{1}$
\\
$^{1}$Department of Physics, University of Oxford, Parks Rd, Oxford, OX1 3PU, UK\\
$^{2}$Department of Physics and Astronomy, University College London, Gower St, London WC1E 6BT, UK\\}

\date{Accepted XXX. Received YYY; in original form ZZZ}

\pubyear{2019}

\begin{document}
\label{firstpage}
\pagerange{\pageref{firstpage}--\pageref{lastpage}}
\maketitle

\begin{abstract}
We present a novel retrieval technique that attempts to model phase curve observations of exoplanets more realistically and reliably, which we call the 2.5-dimension (2.5-D) approach. In our 2.5-D approach we retrieve the vertical temperature profile and mean gaseous abundance of a planet at all longitudes and latitudes \textbf{simultaneously}, assuming that the temperature or composition, $x$, at a particular longitude and latitude $(\Lambda,\Phi)$  is given by \linebreak
$x(\Lambda,\Phi) = \bar{x} + (x(\Lambda,0) - \bar{x})\cos^n\Phi$, where $\bar{x}$ is the mean of the morning and evening terminator values of $x(\Lambda,0)$, and $n$ is an assumed coefficient. We compare our new 2.5-D scheme with the more traditional 1-D approach, which assumes the same temperature profile and gaseous abundances at all points on the visible disc of a planet for each individual phase observation, using a set of synthetic phase curves generated from a GCM-based simulation. We find that our 2.5-D model fits these data more realistically than the 1-D approach, confining the hotter regions of the planet more closely to the dayside. We then apply both models to WASP-43b phase curve observations of HST/WFC3 and Spitzer/IRAC. We find that the dayside of WASP-43b is 
apparently much hotter than the nightside and show that this could be explained by the presence of a thick cloud on the nightside with a cloud top at pressure $< 0.2$ bar. We further show that while the mole fraction of water vapour is reasonably well constrained to $(1-10)\times10^{-4}$, the abundance of CO is very difficult to constrain with these data since it is degenerate with temperature and prone to possible systematic radiometric differences between the HST/WFC3 and Spitzer/IRAC observations. Hence, it is difficult to reliably constrain C/O. \end{abstract}

\begin{keywords}
radiative transfer -- methods: numerical -- planets and satellites: atmospheres -- planets and satellites: individual: WASP-43b
\end{keywords}



\section{Introduction}

Studying the atmospheres of exoplanets is a field that has expanded rapidly in the past decade. Transiting exoplanets have been studied in both primary transit \citep[e.g.,][]{barstow13,barstow2016consistent,evans2018optical,krissansen2018detectability} and secondary eclipse \citep[e.g.,][]{barstow14,lee14,gandhi18}, with each technique providing a unique insight of the  atmospheres of these alien worlds. The primary transit spectrum allows us to probe composition at high altitudes at the day/night terminator of these planets and a number of studies have been made comparing the observed properties and inferred atmospheres of a representative range of planets \citep[e.g.,][]{sing16,barstow2016consistent,pinhas19}, finding a wide range of apparent H$_2$O abundances and obscuration by clouds, with in one study the presence or absence of clouds apparently dependent on the effective temperature of the planet \citep{barstow2016consistent}. Meanwhile the secondary eclipse spectrum allows us to probe the temperature and composition of the dayside. Recently, using the Hubble/WFC3 and Spitzer/IRAC instruments it has become possible to obtain spectroscopic observations of the full phase curve of three exoplanets: WASP-43b \citep[e.g.,][]{stevenson14, mendonca18}, discovered by \cite{hellier11}, WASP-103b \citep{kreidberg18}, discovered by \cite{gillon14} and WASP-18b \citep[e.g.,][]{maxted13, arcangeli19}, discovered by \cite{hellier09}. The HST/WFC3 observations cover wavelengths from $\sim$ 1 -- 2 $\mu$m at a spectral resolution of 
$\Delta\lambda$ $\sim$~0.035~$\mu$m, while the broad Spitzer/IRAC channel observations at 3.6 and 4.5~$\mu$m, which have higher signal-to-noise ratio (SNR) but lower spectral resolution ($\Delta\lambda$ $\sim$ 1 $\mu$m), extend the coverage to higher altitudes with different gaseous sensitivities. Techniques have been developed to invert such phase curves to derive longitudinal brightness temperature maps at individual wavelengths \citep[e.g.,][]{cowan08, burrows08}, but to perform full atmospheric `retrievals' of such data, where we attempt to find a global distribution of temperature and abundance that fits the measured phase curves at all wavelengths \textbf{simultaneously}, the prevalent current technique has been to fit the observed spectra from each phase individually assuming a model that has the same temperature profile and gaseous abundances at all points on the visible disc; this process is then repeated for all the other phases of the planet's orbit. Such a "1-D" model approach ignores the fact that atmospheric conditions will vary across the face of the observable disc and also fails to utilise the constraints on the atmospheric structure arising from observations at nearby phases. A study of former effect has been made by \citet{blecic17} who used a retrieval model to retrieve a single atmospheric temperature profile from a secondary eclipse spectra generated by a GCM simulation of HD189733b. Meanwhile, \citet{Feng16} explored how the assumption of a single 1-D thermal profile can bias the interpretation of synthetic thermal emission spectra generated from a model atmosphere composed of two different thermal profiles (to approximate the appearance of a Hot Jupiter with a sub-stellar region much hotter than the surrounding background). In some cases it was found that considerable biases could arise.

In this paper, we present a novel technique that performs a more consistent retrieval of the complete phase curve to extract atmospheric structure and composition at all latitudes and longitudes simultaneously, which we believe represents a considerable improvement over the traditional 1-D approach. In Section~2 we review the numerical process of simulating disc-averaged spectra and propose an efficient scheme for computing this numerically and in Section~3 we introduce our new 2.5-D retrieval scheme. Section~4 describes the radiative transfer model we use in our retrieval model and Section~5 describes how we validated our retrieval model with synthetic phase curve spectra generated from a model based on a General Circulation Model (GCM) simulation. In Section~6 we apply our new scheme to observed phase curve spectra of \citet{stevenson17} of WASP-43b and we discuss our findings and present our conclusions in Sections 7 and 8.

\section{Calculating disc-averaged spectra from inhomogeneous atmospheres}

When calculating the disc-averaged spectra of planets there are several numerical techniques that can be used, which have various levels of precision. In this work, where we attempt to retrieve observations from disc-averaged spectra of inhomogeneous atmospheres, it was vital that the disc-averaging was treated as accurately as possible, but also that we used a scheme that was not too computationally expensive; this is important since in a retrieval we have to iterate over many possible solutions before we arrive at our best estimate. 

\begin{figure}
\includegraphics[width=\columnwidth]{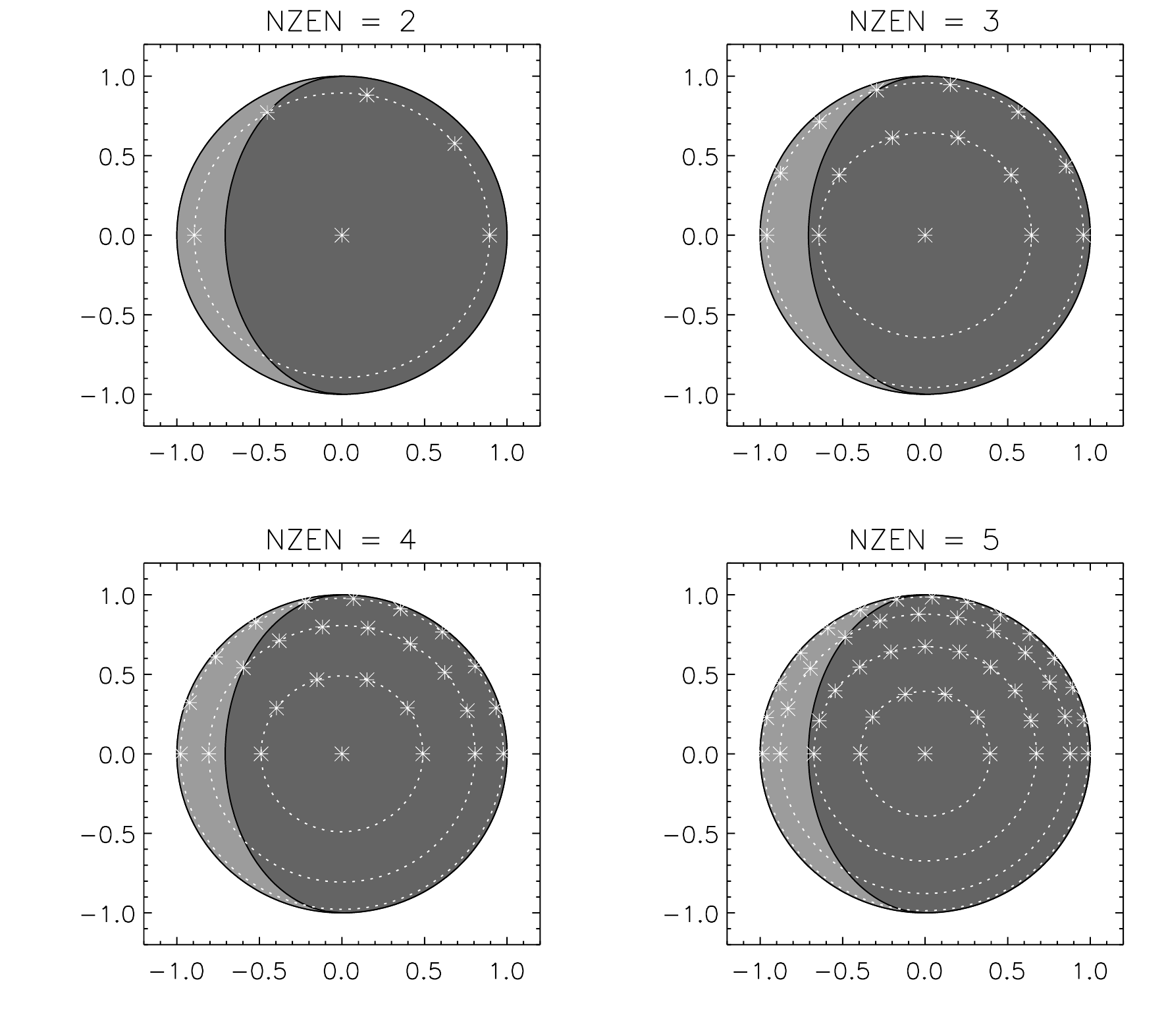}
\caption{Example of disc-averaging quadrature points chosen for $N_{\mu}$ = 2, 3, 4 and 5 and phase angle $\zeta = 45^\circ$ (where $0^\circ$ is nightside and $180^\circ$ is dayside). The daylight side of the planet is coloured light grey, while the nightside is dark grey. The zenith angles of the chosen Gauss-Lobatto quadrature scheme are shown by the dotted white lines. The azimuth points along each zenith angle line (indicated by the asterisk symbols) are chosen to coincide with the terminator for zenith angles that intersect the terminator and be roughly equally spaced with a fractional disc distance of $1.0/N_{\mu}$. As we assume north-south symmetry, only points on or above the equator are needed for disc-integration.\label{fovwt}}
\end{figure}

At the most basic level, the contribution to the total spectral irradiance, $I(\lambda)$ (units of W m$^{-2}$ $\mu$m$^{-1}$),  at wavelength $\lambda$,  seen from an area on the `surface' of a planet  is given by:

\begin {equation} 
dI(\lambda) =  R(\lambda)d\Omega,
\end {equation}

where $R(\lambda)$ is the spectral radiance (W m$^{-2}$ sr$^{-1}$ $\mu$m$^{-1}$) calculated for a particular area on the disc and $d\Omega$ is the solid angle it projects. An element of area $dA$ on the `surface' of a planet at a distance $D$ from the Earth, viewed at a  local zenith angle $\theta$, subtends a solid angle $d\Omega = dA \cos \theta / D^2$ and since for a planet with radius $a$, $dA = a^2 \sin \theta d\theta d\phi$ (where $\theta$ is the local zenith angle, and $\phi$ is the azimuth angle), the total planetary spectral irradiance or  `flux' seen is:

\begin {equation} \label{eq:gen}
F_{plan}  =  \frac{a^2}{D^2}  \int_{\phi = 0}^{2 \pi}  \int_{\theta = 0}^{\pi/2 } R(\lambda, \theta, \phi) \sin\theta\cos\theta d\theta d\phi 
\end {equation}

or 

\begin {equation}  \label{eq:wt}
F_{plan}  = \frac{a^2}{2 D^2} \int_{\phi = 0}^{2 \pi}  \int_{\theta = 0}^{\pi/2 } R(\lambda, \theta, \phi) \sin 2\theta d\theta d\phi.
\end {equation}

By substituting $\mu = \cos\theta$, Eq.~\ref{eq:gen}  can be simplified to:

\begin {equation} \label{eq:wtmu}
 F_{plan} =  \frac{a^2}{D^2} \int_{\phi = 0}^{2 \pi}  \int_{\mu = 0}^{1} R(\lambda, \mu, \phi) \mu d\mu d\phi.
\end {equation}

The disc-averaged radiance $\bar{R}(\lambda)$ (W m$^{-2}$ sr$^{-1}$ $\mu$m$^{-1}$)  can be calculated from $F_{plan}$ as

\begin {equation} 
\bar{R}(\lambda) = \frac{F_{plan}}{\Omega_{total}} = \frac{F_{plan}}{\frac{\pi a^2}{D^2} }.
\end {equation}

We can see from Eq.~\ref{eq:wt} that the planetary flux (and thus disc-averaged radiance) is most strongly weighted by the radiance emitted at a zenith angle of 45$^\circ$. Hence to a rough first approximation, the disc-averaged radiance can be estimated as

\begin {equation} 
\bar{R}(\lambda) = \frac{1}{2 \pi} \int_{\phi = 0}^{2 \pi} R(\lambda, \theta = 45^\circ, \phi) d\phi,
\end {equation}

and even more simply, if we assume azimuthal symmetry, as $\bar{R}(\lambda) = R(\lambda, \theta = 45^\circ)$. However, the most accurate estimate of the disc-averaged radiance (especially if $R(\lambda, \mu, \phi)$ varies significantly with position on the disc)  comes from the full integration:

\begin {equation}  \label {eq:av}
\bar{R}(\lambda) = \frac{1}{\pi} \int_{\phi = 0}^{2 \pi}  \int_{\mu = 0}^{1} R(\lambda, \mu, \phi) \mu d\mu d\phi.
\end {equation}

For our integration scheme we integrated Eq.~\ref{eq:av} with respect to $\mu = \cos \theta$ using a Gauss-Lobatto quadrature scheme, while we used a Trapezium rule integration for the azimuth part of the integration, splitting the circle for each zenith angle into equally-spaced points  on the nightside from disc edge to terminator (the distance between the points on a zenith angle circle being $\sim R/N_\mu$) and equally-spaced points on the dayside from terminator to disc edge (Fig.~\ref{fovwt}) with the same approximate separation.  Our computed disc-averaged radiance was thus calculated as

\begin {equation} \label{eq:main}
\bar{R}(\lambda) = 2 \sum_i^{N_\mu} \sum_j^{N_\phi} R(\lambda, \mu_i, \phi_{ij}) \mu_i \Delta \mu_i w_{ij},
\end {equation}

where $\mu_i$ are the Gaussian quadrature points, $\Delta \mu_i$ are the Gaussian weights, $\phi_{ij}$ are the azimuth angles (which are different for different $\mu_i$) and $w_{ij}$ are the azimuth angle trapezium-rule integration weights. To ensure correct normalisation, $\sum_j w_{ij} = 1.0$ and $\sum_i^{N_\mu} \Delta \mu_i = 1.0$ (i.e., $\sum_i^{N_\mu} \mu_i \Delta \mu_i = 0.5$). In addition, we assumed that the exoplanet had a transit with low impact parameter such that northern and southern hemispheres were observed at the same emission angle and we also assumed north/south symmetry such that we only had to integrate over half the disc.

An example of the chosen quadrature points for our scheme using $N_{\mu} = $ 2, 3, 4 or 5 zenith angles is shown in Fig.~\ref{fovwt} for a phase of 0.125 (i.e., a phase angle $\zeta = 45^\circ$). Here we can see that as we add more zenith angles, the number of sampled points on the disc is greatly increased. In our scheme if a given zenith angle circle intersects the day-night terminator, one azimuth quadrature point is placed on that terminator and the remaining azimuth points are chosen to be roughly equally spaced on the day and nightside, as noted above. The reason for doing this is that if we were to observe a planet where the reflected starlight is strong, we would expect that contribution to fall rapidly towards the terminator. By making sure we place a quadrature point at the terminator, we can thus ensure a better estimate of the azimuthal average of the radiance at a given zenith angle. Although in this study we have not included reflected starlight in our models, the disc-integration scheme we have defined means that we well be able to seamlessly and accurately incorporate this in future. 

\subsection{Computation of contribution of reflected starlight}

Assuming a Lambertian reflecting surface, the scattered radiance at any point on the disc is equal to $\alpha F_{star}/\pi$, where $\alpha$ is the reflectivity and $F_{star}$ is the local stellar irradiance (i.e., W m$^{-2}$), calculated as $F_{star} = P \cos \theta /(4 \pi d^2)$, where $P$ is the total power of the star, $d$ is the distance of the planet from the star, and $\theta$ is the local stellar zenith angle at the point on the disc. At a phase angle of $\zeta = 180^\circ$ the stellar and viewing zenith angles are the same and hence, after Eq.~\ref{eq:wtmu}, the planet flux is: 

\begin {equation} 
F_{plan}  =  \frac{\alpha P a^2}{4 \pi d^2 D^2}  \int_{\phi = 0}^{2 \pi}  \int_{\theta = 0}^{\pi/2 } \sin\theta \cos ^2\theta d\theta d\phi 
\end {equation}

or 

 \begin {equation} 
F_{plan}  =  \frac{\alpha P a^2}{2 d^2 D^2}  \int_{0}^{1} \mu^2 d\mu = \frac{\alpha P a^2}{6 d^2 D^2}.  
\end {equation}

Hence, since $F_{star} = P/(4 \pi D^2)$  we have

\begin {equation} \label{eq:refl}
\frac{F_{plan}}{F_{star}}  = \alpha \frac{2}{3} \frac{a^2}{d^2}  
\end {equation}

which, at a general phase angle $\zeta$ can be approximated to

\begin {equation} 
\frac{F_{plan}}{F_{star}}  = \alpha \frac{2}{3} \frac{a^2}{d^2}\frac{1-\cos \zeta}{2}.
\end {equation}

We will return to the likely contribution of reflected starlight to the observed phase curve observations of WASP-43b later.

\begin{figure*}
\includegraphics[width=\textwidth]{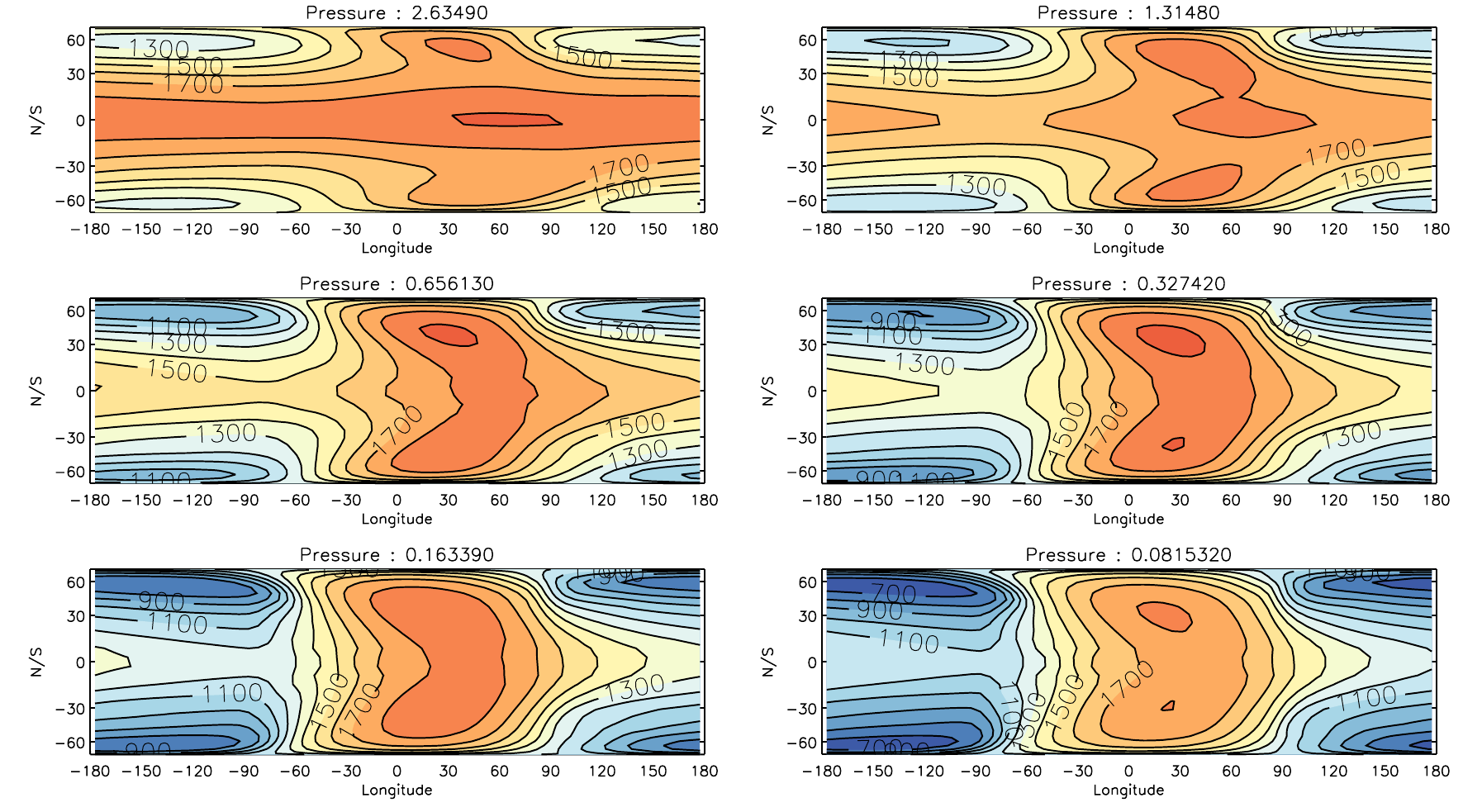}
\caption{Temperature latitude/longitude contour maps generated by our GCM-based model at six different pressure levels. \label{GCMA}}
\end{figure*}

\section{Comparison of 1-D and 2.5-D Retrieval Schemes}

The simplest way of retrieving atmospheric information from phase curve observations, such as those reported by \citet{stevenson17}, is for each phase angle, $\zeta$, in the set of $N_{phase}$ phase angles to assume that the hemisphere facing Earth has the same temperature and gaseous abundance profiles at all visible parts of the disc. A disc-averaged spectrum can then be computed very simply (in the most simple cases by just computing the spectrum at a viewing zenith angle of $45^\circ$) and the temperature/abundance profiles retrieved using schemes such as optimal estimation \citep[e.g.,][]{rodgers00,irwin08} or more generally with Bayesian methods, such as nested sampling \citep[e.g.,][]{skilling06}. Such techniques, which we will here refer to as one-dimensional (i.e., 1-D) retrievals, are easy to understand and implement, but suffer from a crucial systematic disadvantage: in order to model the disc-averaged spectra we assume homogeneous conditions across the visible disc, but we then use these retrievals to infer that the atmospheric conditions change markedly with the sub-stellar longitude. Hence, for observations at phase angles sampling mostly the nightside, for example, we assume we cannot see the dayside at all, which is demonstrably incorrect, geometrically, as can be seen in Fig.~\ref{fovwt}. Here, we aimed to develop a technique where the spatial variation of atmospheric conditions with position on the planet was incorporated directly within the radiative transfer scheme used to retrieve them.   

In our new retrieval scheme, for a set of $N_{phase}$ phase curve spectral observations, we fitted all $N_{phase}$  spectra simultaneously with a model that contained the temperature profile and mean gaseous abundances at $N_l$ equally-spaced longitudes covering 180$^{\circ}$W to 180$^{\circ}$E. We then set the value of either temperature or abundance, $x(\Lambda,\Phi)$, at a longitude and latitude $(\Lambda,\Phi)$ to be

\begin{equation} \label{eq:n}
x(\Lambda,\Phi) = \bar{x} + (x(\Lambda,0) - \bar{x})\cos^n\Phi,
\end{equation}

where $x(\Lambda,0)$ is the modelled value of the abundance or temperature (at a particular pressure level) at the equator and longitude, $\Lambda$, and $\bar{x}$ is mean of the morning and evening terminator values of $x(\Lambda,0)$, i.e., $\bar{x} = (x(-90^\circ,0) + x(90^\circ,0))/2$. For the exponent of $\cos \Phi$ in this equation, $n$, we used values of $n=0.25$ (assuming Stefan-Boltzmann-like equilibrium with the stellar irradiation), but also values such as $n=1$ and $n=2$ to test for likely latitudinal variation. With this parameterisation we can then calculate the disc-averaged radiance over an inhomogenous disc using our numerical disc-integration scheme described above, but more importantly we can retrieve directly the atmospheric conditions on a planet as a function of longitude and height (by retrieving the set of values $x(\Lambda,0)$) and also have some sensitivity to latitudinal distribution through the assumed $\cos^n \Phi$ dependence. Hence, we call our new technique a 2.5-dimensional (i.e., 2.5-D) retrieval scheme. 

Since we aim to retrieve both temperature and abundances from disc-averaged spectra where we have to perform radiative transfer calculations at a large number of positions across the visible disc for each of the $N_{phase}$ phase angles, our radiative-transfer, or `forward model', is computationally intensive. Hence, for this paper we have used the retrieval method of optimal estimation \citep[e.g.,][]{rodgers00}. Although less suitable than Bayesian approaches in cases of observations with low signal-to-noise ratio (SNR), where we need to assess the suitability of a wide range of solutions, the optimal estimation method is very much faster and was thus useful in setting up and testing this approach. For future work, however, we recommend implementing this approach within a Bayesian framework. 

The method of optimal estimation is not used as often for exoplanetary work as Bayesian methods \citep[although it has been used several times before, e.g.,][]{lee14,barstow2016consistent} and some aspects of it may appear counter-intuitive to readers more familiar with techniques such as Monte-Carlo Markov Chain (MCMC) analysis and Nested Sampling. The fundamental idea of the approach is that we start with an \textit{a priori} estimate of the atmospheric state and associated \textit{a priori} errors (contained in an error covariance matrix that holds, for example, the expected variances of the temperature profile as a function of height and longitude, and the covariances between these temperatures) and then find a solution that most closely matches the observations without deviating too greatly from the \textit{a priori} estimate by minimising the cost function $\phi$:

\begin {equation} \label{eq:phi}
\phi = {(\mathbf{y}_m - \mathbf{y}_n)}^\mathrm{T} \mathbf{S}_\epsilon^{-1} (\mathbf{y}_m - \mathbf{y}_n) + 
{(\mathbf{x}_n - \mathbf{x}_0)}^\mathrm{T} \mathbf{S}^{-1} (\mathbf{x}_n - \mathbf{x}_0),
\end {equation} 

where $\mathbf{y}_m$ is the measurement vector, composed of the measured observations, $\mathbf{y}_n$ is the vector of modelled observations calculated from the state vector of model parameters $\mathbf{x}_n$, $\mathbf{S}_\epsilon$ is the measurement covariance matrix, $\mathbf{S}$ is the \textit{a priori} covariance matrix and $\mathbf{x}_0$ is the \textit{a priori} state vector. The approach \citep[e.g.,][]{rodgers00} was originally developed for analysing Earth observations, where we have very good \textit{a priori} estimates of the expected atmospheric state from climatology and \textit{in situ} observations. However, in its formulation the \textit{a priori} error covariance matrix effectively provides a means of `braking' or `smoothing' the solutions, with small \textit{a priori} error values leading to solutions that differ little from the \textit{a priori} estimates and large values leading to solutions that fit the observed data closely, but are not constrained by the \textit{a priori} estimates at all. The problem with the latter type of solution, known as the "exact" solution, is that it is "ill-conditioned": random noise in the observations can potentially lead to the appearance of large amplitude oscillations in the retrieved vertical profiles. Although such profiles may generate synthetic spectra that match the observations very well, we do not, from atmospheric physics modelling experience, expect the retrieved temperature profiles to vary very rapidly with height. For planetary work, where we do not have a well known \textit{a priori} based on climatology, models such as our NEMESIS code \citep{irwin08} instead use the \textit{a priori} errors as adjustable `tuneable' parameters to vary the balance of the retrieval between not differing too far from some \textit{a priori} estimate and fitting the observations so closely that the retrieval becomes ill-conditioned. In essence, reducing the \textit{a priori} temperature error in such a scheme reduces the ability of the model to `bend' the fitted temperature profile too greatly from the \textit{a priori} temperature profile to generate spectra that better fit those observed. The diagonal elements of the \textit{a priori} covariance matrix $\mathbf{S}$ are set to the square of the \textit{a priori} error for each state vector element, while to achieve the vertical and horizontal smoothing required by our 2.5-D model the off-diagonal elements are set \citep[e.g.,][]{irwin08} as 

\begin {equation} 
S_{ij}  =  (S_{ii} S_{jj})^{1/2} \exp \bigg(-\frac{|i-j|}{l}\bigg), 
\end {equation}

where $l$ is a `correlation length' that adjusts the degree of covariance and thus smoothing. We will return to the specific settings used in our model later.


\section{Radiative transfer model}

In this paper our aim was to model the phase curve observations of WASP-43b, reported by \citet{stevenson17}, using our NEMESIS retrieval model \citep{irwin08} in its traditional optimal estimation mode using either the 1-D approach or our new 2.5-D approach. The measured phase curves of \citet{stevenson17} include 15 curves extracted from HST/WFC3 observations, binned in equally-spaced bins of width 0.035 $\mu$m covering the wavelength range 1.1425 -- 1.6325 $\mu$m, and two phase curves from Spitzer/IRAC from broad channels (see Fig.~\ref{summary}) centred at 3.6 and 4.5 $\mu$m. Although \citet{mendonca18} and \citet{morello19} have independently reprocessed the Spitzer/IRAC observations and found significantly different fluxes at 3.6 and 4.5 $\mu$m, we have used the measured phase curves of \citet{stevenson17} here for ease of comparison of our results. We modelled the spectra using the method of correlated-k \citep{lacisoinas91}, using k-distribution look-up tables generated from the most recently available line data \citep[e.g.,][]{garland19}, in particular H$_2$O \citep{barber06}, CH$_4$\citep{yurchenko14}, CO$_2$ \citep{tashkun11} and CO \citep{rothman10}. For high-resolution calculations we used these k-tables at their native resolution of 0.02 $\mu$m. However, for our retrievals, where we needed to compute the spectra for multiple iterations, we calculated averaged k-tables for each of the 17 individual `channels'. Test calculations proved this to be an excellent approximation, even for broad Spitzer/IRAC channels, resulting in residuals far less than the quoted measurement noise of \citet{stevenson17}. The k-tables were compiled separately for each gas and combined during the radiative-transfer calculation using the random-overlapping-line approximation \citep{lacisoinas91}. Again, we found this approximation to give residuals insignificant compared with measurement error, as has been found by other authors \citep[e.g.,][]{molliere15}. Collision-induced absorption of H$_2$--H$_2$ and H$_2$--He was modelled using the coefficients of \cite{borysow1989a} and \cite{borysow1989b}. Since in this study we assumed thermal emission only, we were able to use NEMESIS in its implicit differentiation mode, where the functional derivatives (i.e., rate of change of computed radiance with atmospheric parameter) are calculated implicitly and thus very rapidly. The planet was modelled to have a mass of 2.052 M$_J$ and radius 1.036 R$_J$ \citep{gillon2012trappist}, orbiting its star at a distance of 0.015AU. The planet orbits a main sequence K-type star with a mass of 0.717M$_\odot$, radius of 0.667 R$_\odot$, effective temperature of 4520 $\pm$ 120K and metallicity [Fe/H] = -0.01 dex \citep{gillon2012trappist}. These parameters were used to generate a stellar spectrum from the Kurucz ATLAS model atmospheres \citep{castelli2004new}. This was compared with a spectrum extracted for the same star from PHOENIX \citep{allard00}. While there were some small differences, the effect on the retrieved atmospheric parameters was found to be small.

\section{Validation of retrieval schemes with synthetic observations}

\begin{figure}
\includegraphics[width=\columnwidth]{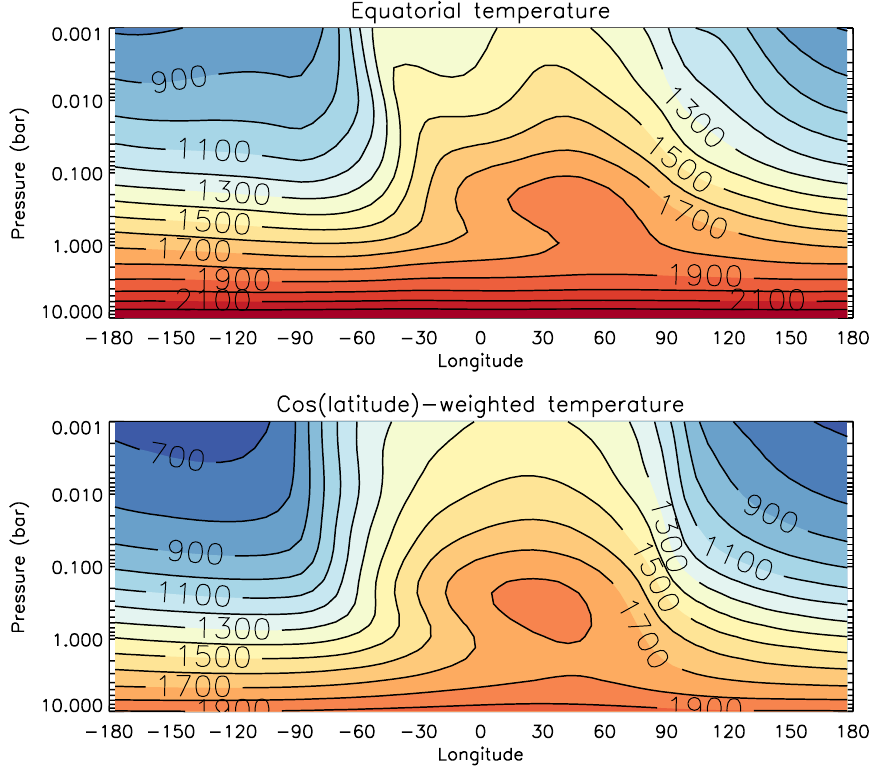}
\caption{Averaged temperatures and abundances in GCM-based model. Top panel shows the temperature at the equator as a function of longitude and pressure. Bottom panel shows the average temperature profile for each longitude, averaged over all latitudes using cos(latitude) as a weight. \label{GCMB}}
\end{figure}

\begin{figure*}
\includegraphics[width=\textwidth]{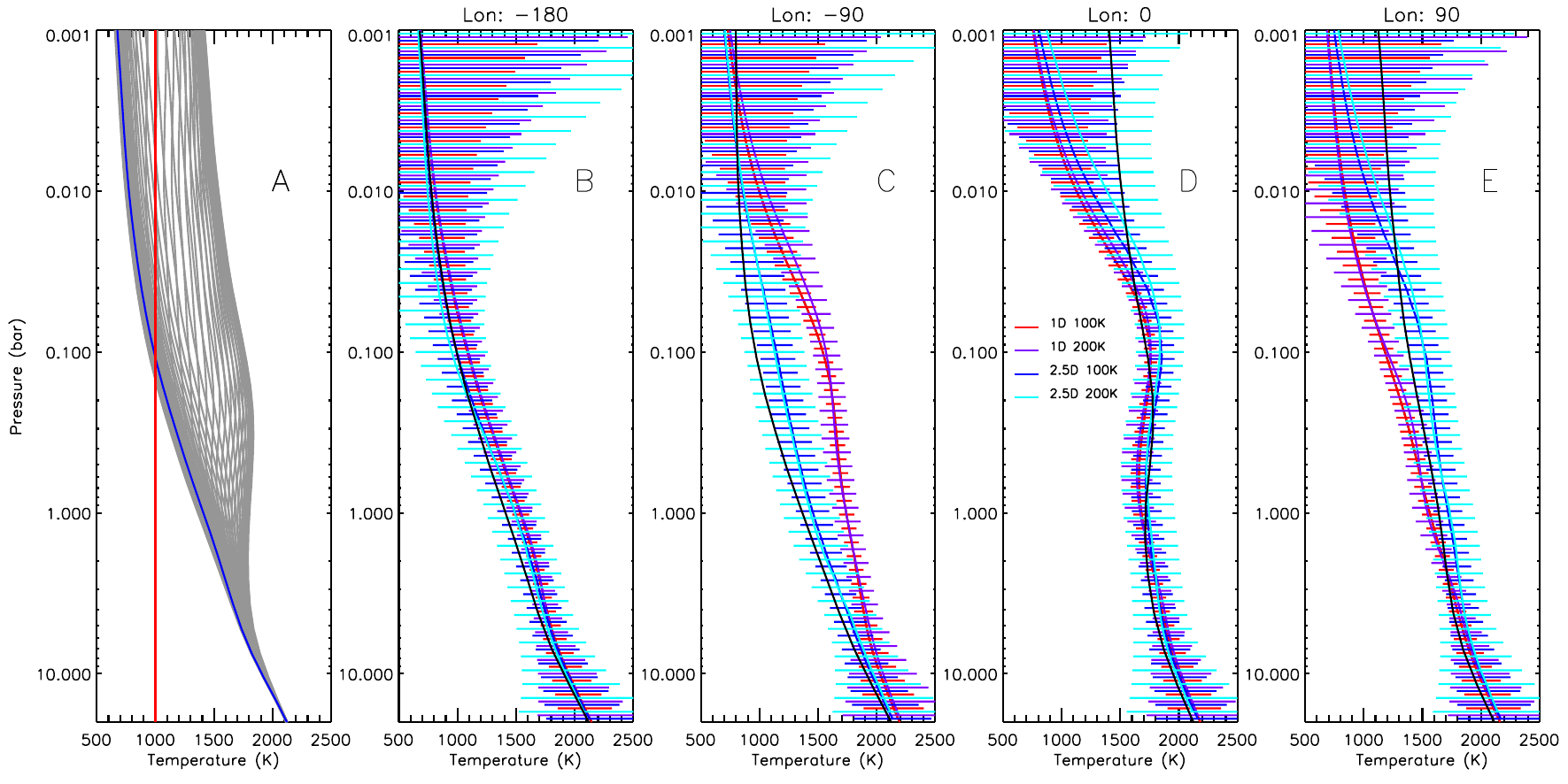}
\caption{Panel A: Latitudinally-averaged temperature profiles for all longitudes from our GCM-based model (using cos(latitude) as a weight), compared with the two \textit{a priori} temperature profiles used in our study, one set to $1000$~K at all levels (red), and the other using the latitudinally-averaged nightside temperature profile at a solar longitude of 180$^\circ$. The \textit{a priori} errors were set to either $\pm 50$, $\pm 100$, or $\pm 200$~K at all levels. Panels B -- E: Comparison of temperature profiles retrieved from GCM-based synthetic phase curve spectra at four different central meridian longitudes using our 1-D and 2.5-D retrieval models: --180$^\circ$, --90$^\circ$, 0$^\circ$ and 90$^\circ$. The horizontal bars indicate the error due to measurement noise, while the black profile indicates the `true' latitudinally-averaged GCM-based temperature profiles at these longitudes. Since the retrieved error, $\sigma_r$, tend back to \textit{a priori} error, $\sigma_a$, where there is little information, we have here plotted the equivalent raw measurement error $\sigma_m = \sqrt(1/\sigma_r^2 - 1/\sigma_a^2)$, which better shows the unconstrained retrieval error.  \label{GCMC}}
\end{figure*}

Before applying our model to the observed WASP-43b phase curve spectra of \citet{stevenson17} themselves, we needed to test both the 1-D and 2.5-D schemes for accuracy and reliability. To do this we used a global thermal structure calculated for WASP-43b from a global circulation model (GCM) SPARC/MITgcm \citep{showman09}. The specific model for WASP-43b was run based on the setup of \citet{parmentier16}
and produces results extremely similar to the published model of \cite{kataria15}.  The model we used was generated assuming no clouds and the gas abundances varied with altitude and location according to equilibrium chemistry leading to rather significant variation with longitude for CH$_4$. However, because of horizontal `quenching' \citep[e.g.,][]{cooper06,agundez14} we do not expect there to be significant longitudinal variation in the abundance of any gas. Hence, to enable a simpler test of our retrieval approaches, the gas abundances for H$_2$O, CO, CO$_2$ and CH$_4$ were reset at all altitudes and locations to be the latitudinally-averaged abundances in the 0.1 -- 1-bar pressure region (using $\cos\Phi$ as the weight, where $\Phi$ is the latitude) at the sub-stellar meridian. This gave abundances for H$_2$O, CO$_2$, CO and CH$_4$ of $4.8\times 10^{-4}$, $7.4 \times 10^{-8}$, $4.6\times 10^{-4}$ and $1.3\times 10^{-7}$, respectively.  The modified abundance profiles at all latitudes and longitudes were then used to generate a set of synthetic  HST/Spitzer $F_{plan}/F_{star}$ spectra that we could then use as our `ground-truth' in our retrieval tests. Although the approach is not fully self-consistent \citep[varying the abundances and thus atmospheric gas opacities will affect the modelled thermal structure also in a self-consistent scheme, see e.g.,][]{Drummond2018,Steinrueck2019}, this is not a concern for the purpose of testing our retrieval schemes since here we retrieve the temperature and abundance profiles independently, with no assumption on the radiative equilibrium state of the atmosphere. Figure~\ref{GCMA} shows the horizontal temperature distribution at six different pressure levels in this model, where we can see that the sub-stellar hot spot is shifted eastwards at all levels, but less so at mid-latitudes and lower pressures. We have plotted these contour maps as a function of longitude and latitude, with the plotted latitudinal distance $\Phi$ weighted by $\cos\Phi$ to mimic the fact that we observe WASP-43b from above the equator and thus polar latitudes are foreshortened. Figure~\ref{GCMB} compares the longitudinal variation of the equatorial temperature profile with a latitudinally-averaged temperature profile, where the weights are set as $\cos\Phi$. Such an average can be seen to smooth over much of the structure seen in the equatorial temperature profiles, more tightly constrains the hotter regions of the planet to the dayside (i.e., between longitudes of $-90^\circ$ and $90^\circ$E), and also moves the centre of the hot spot slightly westwards. Finally, Fig.~\ref{GCMC} compares the latitudinally-averaged temperature profiles for all longitudes together with test retrieved profiles that are described later. 

\begin{figure*}
\includegraphics[width=\textwidth]{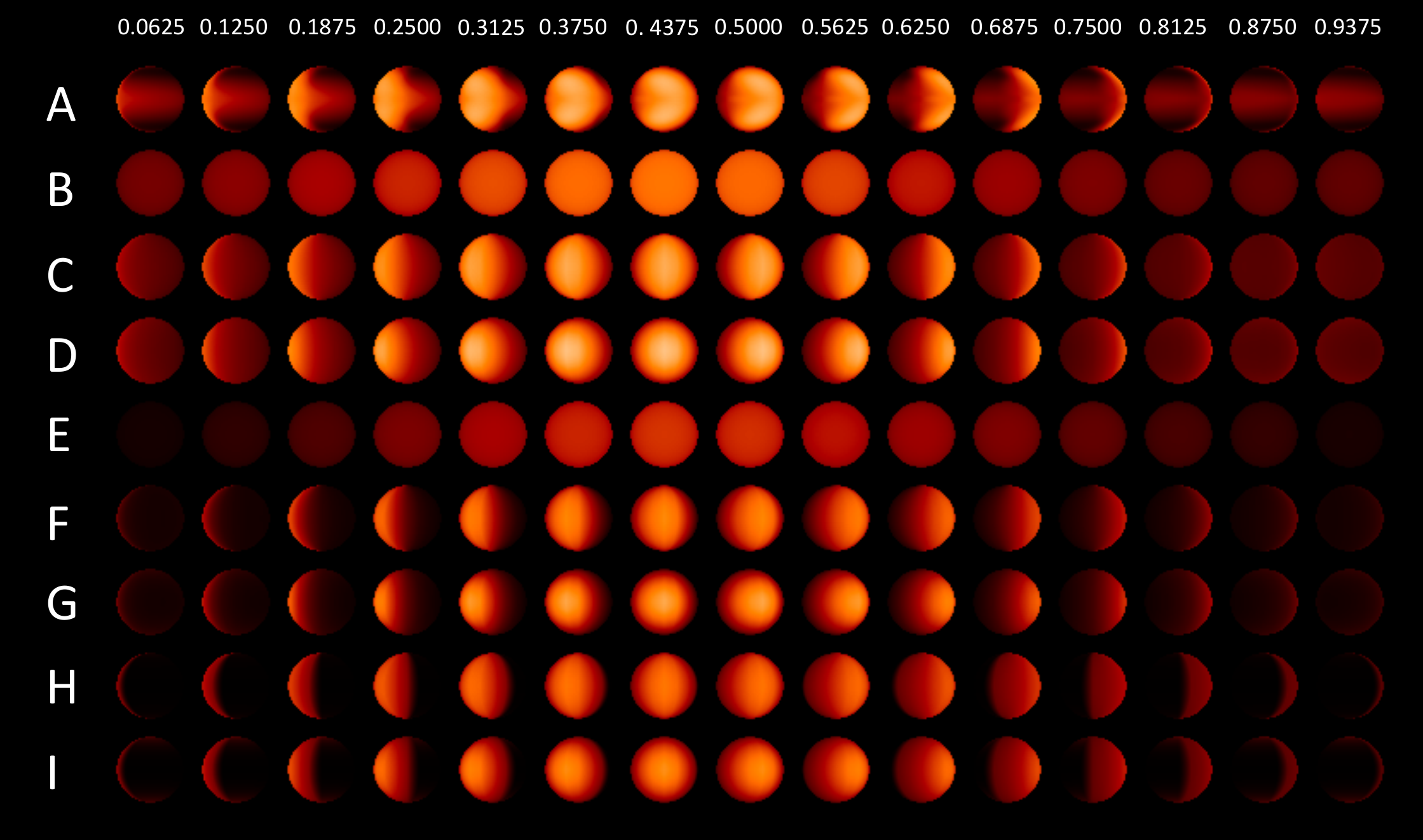}
\caption{Simulated appearance of disc of WASP-43b at 4.5 $\mu$m at different phases for nine different cases: A) The GCM-based model atmosphere; B) Retrieved appearance from GCM-based synthetic phase curves with 1-D model; C) Retrieved appearance from GCM-based synthetic phase curves with 2.5-D model, assuming $(\cos \Phi )^{0.25}$ latitude dependence; D) Retrieved appearance from GCM-based syntheticl phase curves with 2.5-D model, assuming $\cos \Phi$ latitude dependence; E) Retrieved appearance from real observed phase curves with 1-D model; F) Retrieved appearance from real observed phase curves with 2.5-D model, assuming $(\cos \Phi )^{0.25}$ latitude dependence; G) Retrieved appearance from real observed phase curves with 2.5-D model, assuming $\cos \Phi$ latitude dependence; H) Retrieved appearance from real observed phase curves with 2.5-D model, assuming $(\cos \Phi )^{0.25}$ latitude dependence and forcing nightside temperatures to 500~K at all altitudes; and I) Retrieved appearance from real observed phase curves with 2.5-D model, assuming $\cos \Phi$ latitude dependence and forcing nightside temperatures to 500~K at all altitudes. The phase angle increases from left to right in all cases and is indicated at the top of the figure.\label{phasim}}
\end{figure*}

The phase curve observations of WASP-43b of \citet{stevenson17} give the $F_{plan}/F_{star}$ ratios at fifteen different phases between 0.0625 and 0.9375. Hence, we computed the spectra we would expect to see with our model WASP-43b atmosphere at these same phases, assuming thermal emission only.  We computed synthetic phase curve spectra using our disc-integration scheme for $N_\mu$ = 2, 3, 4 and 5 zenith angles and found that the computed phase curves converged rapidly as $N_\mu$ increased, with $\chi^2/n$ (where $n$ is the number of spectral points) reducing by 0.4 going from $N_\mu$ = 2 to 3, further falling by 0.02 going from $N_\mu$ = 3 to 4 and decreasing by only 0.003 going from $N_\mu$ = 4 to 5 (using the measurements errors of \cite{stevenson17} for computing $\chi^2/n$). These differences are negligible compared with measurement error for $N_\mu > 2$, but to be sure of numerical accuracy we used  $N_\mu = 5$ in all further calculations. As can be seen later in Figs.~\ref{comparephase} and \ref{comparephase_qt} our model atmosphere provides a reasonable first approximation to the observed phase curves and thus we were satisfied that the GCM-based model provided a reasonably representative data set with which to test our retrieval method. In addition to the disc-averaged spectra, we also generated from the model atmosphere images of the disc of WASP-43b as it would appear at the WFC3/IRAC wavelengths for each phase, i.e., how the planet would appear to an observer who was able to spatially resolve the planet. These images, for the 4.5-$\mu$m Spitzer/IRAC channel, are shown in Fig.~\ref{phasim}. In addition to this primary set of synthetic phase curves, we also generated a secondary set where randomly generated Gaussian noise was added to the spectra with the standard deviation set to the quoted error estimates of \citet{stevenson17}. 

We tested our retrieval model with two \textit{a priori} temperature profiles:  one where the \textit{a priori} temperature profile at all longitudes was set to the latitudinally-averaged  nightside temperature profile of our model at $180^\circ$E (using $\cos\Phi$ as a weight); and 2) one where the \textit{a priori} temperature profile was set to a constant temperature of 1000~K at all altitudes and longitudes. In both cases we found in `tuning' tests that using \textit{a priori} errors of either $\pm 50$, $\pm 100$, or $\pm 200$~K at all altitudes, with the vertical correlation length in the \textit{a priori} covariance matrix set to $\Delta H$ = 1.5 pressure scale heights, allowed the retrieval model to fit to the observed spectra reasonably well, but were not so large that the retrieval model became `ill-conditioned' (as described in Section 3).  These profiles are shown for reference in Fig.~\ref{GCMC}. It is important to recognise that these assumed \textit{a priori} errors are not hard limits as they might be in models such as MCMC or Nested Sampling, but instead set the relevant weights in the cost function (Eq. \ref{eq:phi}) between fitting the observations well without differing too greatly from the \textit{a priori}. If the data are sufficiently constraining, the fitted temperatures can differ from \textit{a priori} by much more than the  \textit{a priori} error. Using either \textit{a priori} temperature profile we found that our retrieval model tended smoothly to  consistent temperature solutions at pressures where we have most sensitivity ($\sim 0.1 - 1$ bar), and tended smoothly back to \textit{a priori} values at pressures outside this range. However, although retrievals with both \textit{a priori} profiles were consistent, we found the former profile resulted in better behaved retrievals when applied to the spectra measured at all phases since the temperature profile remained close to a physically plausible solution on the nightside, were the SNR is low, and led to more realistic-looking retrievals on the dayside with temperatures tending back to the \textit{a priori} of a smooth increase with pressure at pressures $> 1$ bar, consistent with expectations. Hence, this was our preferred \textit{a priori} profile in the results we describe.
 
Our atmosphere had an assumed He/H$_2$ mole fraction ratio of 0.1765 and contained four additional gases: H$_2$O, CO$_2$, CO and CH$_4$, whose abundances were fitted or assumed. The gas abundances were assumed to have a constant mole fraction (or volume mixing ratio, VMR) with height and when allowing abundance to vary we retrieved logarithmic scaling factors, i.e., VMR = VMR$_{apriori} \exp(x)$, where $x$ is the relevant state vector element, to ensure the mole fractions remained positive during the retrievals. At each iteration the H$_2$ and He abundances were adjusted (keeping He/H$_2$ fixed) to ensure the sum of mole fractions added to unity. The mean molecular weight of the atmosphere was then recomputed in order to calculate the scale height accurately and thus determine the altitudes of each level (above the 20 bar level) assuming hydrostatic equilibrium.

\begin{figure*}
\includegraphics[width=\textwidth]{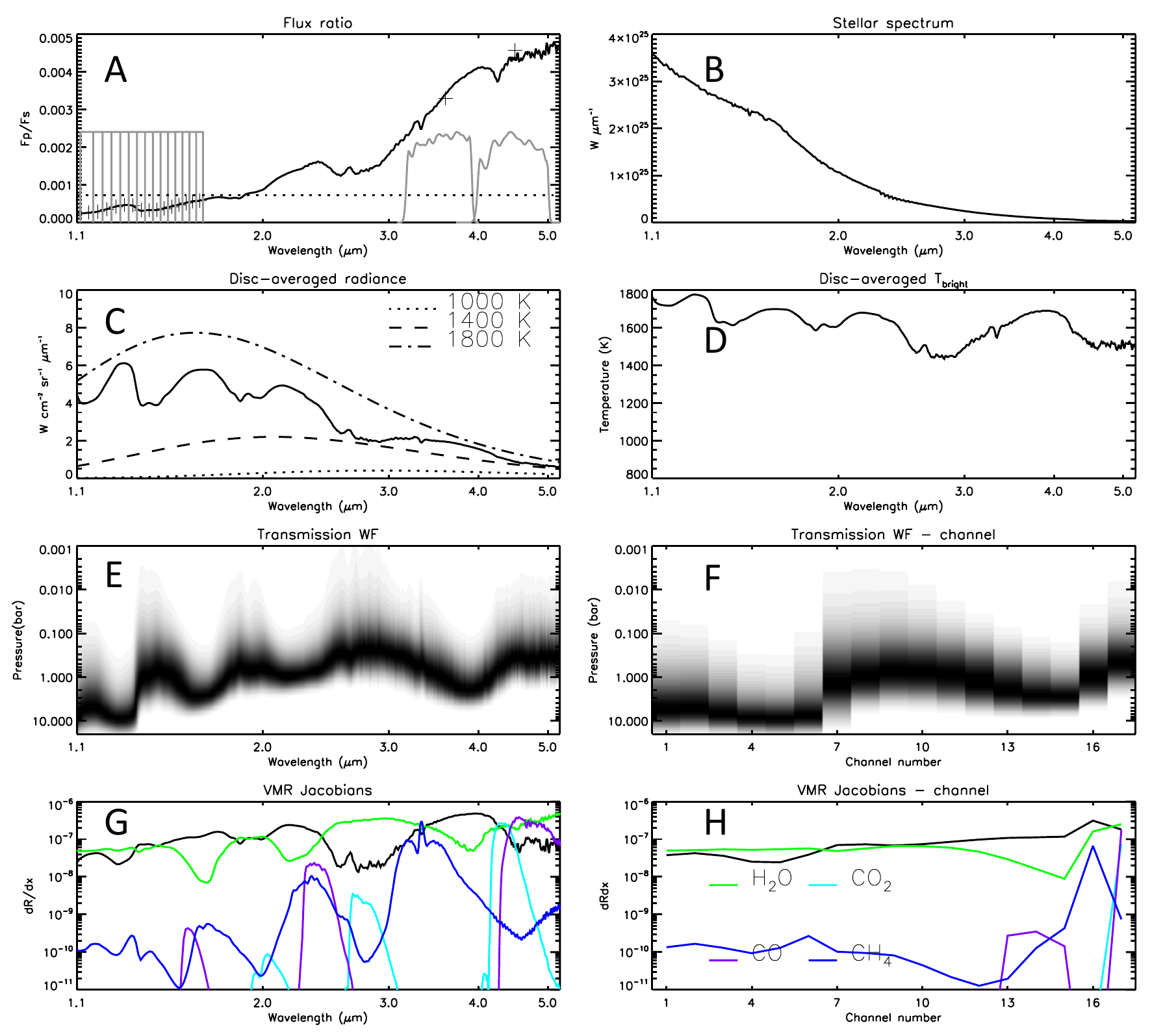}
\caption{1-D model fit to synthetic data generated from our GCM-based model for a phase of 0.5 (i.e., phase angle $\zeta = 180^\circ$). A) modelled $F_{plan}/F_{star}$ spectrum at a resolution of 0.02 $\mu$m using our $N_\mu =5$ quadrature scheme. The grey lines indicate the WFC3/Spitzer filter functions, while the horizontal dotted line shows the maximum possible estimated reflected stellar contribution (assuming unit reflectivity, although since the albedo of WASP-43b is estimated to be $\sim 0.24$ the likely reflected component is considerably less). B) Assumed stellar power spectrum. C) Modelled disc-averaged radiance spectrum, compared with Planck functions for different temperatures. D) Modelled disc-averaged spectrum plotted as equivalent brightness temperature. E) Modelled transmission weighting functions (normalised to unity at each wavelength) at high resolution (i.e., 0.02 $\mu$m) plotted against wavelength. F) Normalised modelled transmission weighting functions integrated over filter function and plotted against channel number. G) Jacobians (plotted against wavelength) for gas abundances (coloured lines - multiplied by 1000) compared with Jacobian for temperature at a $p \sim 1$ bar (black line). H) As Panel G, but channel-averaged values plotted against channel number; a key for the gas colour lines is also included.   \label{summary}}
\end{figure*}

\begin{figure*}
\includegraphics[width=\textwidth]{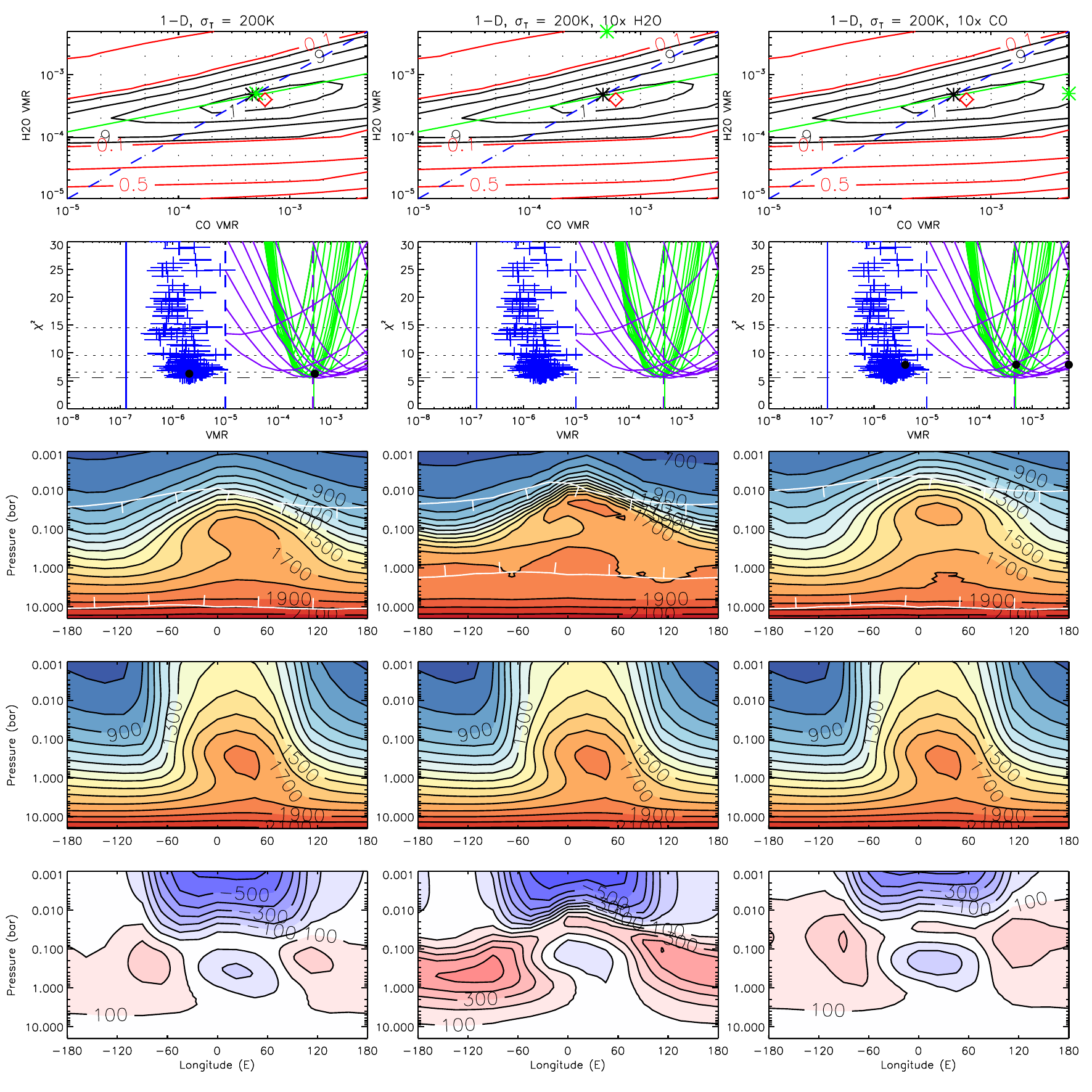}
\caption{Retrieval of GCM-based synthetic phase curve data (excluding randomly generated synthetic noise) using our 1-D model. Figure shows three sets of retrievals: Column 1) 1-D retrieval with the CO and H$_2$O abundance for the temperature plots set to the GCM-based `true' values of $\sim 5 \times 10^{-4}$; Column 2) 1-D retrieval with H$_2$O abundance for the temperature plots increased to  $5 \times 10^{-3}$; and Column 3) 1-D retrieval with CO abundance for the temperature plots increased to  $5 \times 10^{-3}$. The top row shows contour plots of the total $\chi^2$, with contours set at min($\chi^2$), min($\chi^2$)+1, min($\chi^2$)+4, min($\chi^2$)+9, min($\chi^2$)+16. Red contours indicate $\chi^2/n$. For 3-$\sigma$ detection we should be within min($\chi^2$)+9. The blue dashed lines in this row indicates where the abundance of CO is the same as that of H$_2$O, while the green line indicates the approximate principal axis of the best fitting solutions for CO and H$_2$O. In each plot the black asterisk marks the `true' mole fractions, the red diamond indicates the best fitting mole fractions and the green asterisk indicates the abundances chosen for the temperature plots in the remaining rows. Row 2 shows plots of $\chi^2$ vs H$_2$O (green) and CO (purple) mole fraction. Also indicated are the retrieved abundances of CH$_4$ (blue), with the retrieved errors indicated by the horizontal bars. The `true' GCM-based abundances are indicated by the solid vertical lines using the same colour scheme (N.B., the \textit{a priori} CO and H$_2$O values are effectively indistinguishable for this overall range of abundances at $\sim 5 \times 10^{-4}$, so the H$_2$O abundance is plotted as solid green and the CO abundance overplotted as dashed purple). The dashed vertical blue line is the \textit{a priori} CH$_4$ abundance. The horizontal lines in this row show the minimum $\chi^2$ value (dashed lines) and min($\chi^2$)+1, min($\chi^2$)+4, min($\chi^2$)+9 (dotted lines). Finally, the black filled circles in this row indicate the abundances chosen for the temperature plots in the remaining rows. Row 3: Retrieved temperatures as a function of longitude and pressure. The white contour indicates the region where the retrieval is well constrained by the observations and has higher values of the `Improvement factor', $I$, defined as $I = 1-\sigma_r/\sigma_a$, where $\sigma_a$ is the \textit{a priori} error and $\sigma_r$ is the retrieval error. Row 4: $\cos\Phi$ - weighted latitudinally-averaged GCM-based temperature profiles. Row 5: Retrieved temperature minus latitudinally-averaged GCM-based temperature (red-blue scale, white = zero difference). \label{synthetic_ret1}}
\end{figure*}

For our first test, we took our synthetic disc-averaged spectrum (with no additional noise) for a phase of 0.5 (i.e., phase angle $\zeta = 180^\circ$, or secondary eclipse) and performed a 1-D retrieval assuming thermal emission only. Since we assume the same atmospheric conditions at all locations on the disc in our 1-D scheme, the azimuthal part of the integration in Eq.~\ref{eq:main} was not required and so we just integrated over the $N_\mu=5$ zenith angles. We set our \textit{a priori} scaling factors to $x = 0 \pm 1$, with \textit{a priori} mole fractions set to the averaged dayside abundances of the GCM simulation, namely $4.8\times 10^{-4}$, $7.4 \times 10^{-8}$, $4.6\times 10^{-4}$ and $1.3\times 10^{-7}$, for H$_2$O, CO$_2$, CO and CH$_4$, respectively. We then split the atmosphere into 50 layers from 20 bar to $10^{-5}$ bar (split equally with respect to log(pressure)) and retrieved for temperature profile and gas abundances. Our \textit{a priori} temperature profile here was set to the latitudinally-averaged nightside temperature profile of our GCM at $180^\circ$E, with an error of $\pm 100$~K. Figure \ref{summary} shows the fitted spectral flux ratio spectrum in the seventeen HST/Spitzer `channels' compared with a spectrum recalculated from the fitted atmospheric state at reasonably high resolution ($\Delta \lambda = 0.02 \mu$m) over the whole spectral range, showing the channel response functions of the HST/WFC3 and Spitzer/IRAC observations considered. Panel A of Fig.~\ref{summary} also shows the highest likely reflected solar component, which was calculated from Eq.~\ref{eq:refl} assuming a reflectivity $\alpha = 1$. Comparing this with the observed fluxes we find that at the longer wavelengths the reflected component is likely a relatively small part of the total observed flux. However, it seems that reflected sunlight could potentially be a significant component at the shorter wavelengths. \citet{stevenson14} deduce a Bond albedo of $0.18^{+0.07}_{-0.12}$ from the HST/WFC3 observations and more recently, \cite{Keating17} have reanalysed these data to deduce a near-infrared geometric albedo of $0.24 \pm 0.01$ and note that reflected light may be significant at the WFC3 wavelengths. However, given the fact that the likely estimated component of reflected starlight at the  HST/WFC3 wavelengths is smaller than the estimated error bars on these observations, we have chosen to ignore any reflected component in this analysis, as was also done by \citet{stevenson17}.

Figure \ref{summary} also shows the transmission weighting functions computed by our retrieval model (i.e., $dT/dz$, where $T$ is the transmission to space and $z$ is the altitude), showing that our sensitivity to temperature variations peaks between 10 and 0.05 bar. We can thus use these observations to retrieve the temperature between these levels, but outside this pressure region the temperatures will tend back to \textit{a priori} in our optimal estimation scheme. Figure \ref{summary} also compares the Jacobians for gas abundance (i.e., the rate of change of measured signal with respect to change in abundance) with that of temperature near 1 bar. It can be seen that CO$_2$ absorption is only detectable in the 4.5-$\mu$m IRAC channel, while CO absorption is significant in the 4.5-$\mu$m IRAC channel and the three longer wavelength WFC3 channels. However, since the observation in the  4.5-$\mu$m IRAC channel  has much lower measurement error than the WFC3 channels, we can see that there will likely be a degeneracy in the retrieval between the temperature at $\sim 0.1$ bar and the CO/CO$_2$ abundance. The Jacobian for CH$_4$ absorption is more widely distributed with wavelength, but peaks in the 3.6-$\mu$m IRAC channel, which is again measured to a higher precision than the WFC3 channels. Hence, there will likely be a degeneracy in the retrieval between the temperature at just over $0.1$ bar and the CH$_4$ abundance. It can be seen that the Jacobian for water vapour abundance is almost perfectly anti-correlated with the Jacobian for 1-bar-temperature. This is easily understood since water is the main absorber and has features across the 1 -- 5 $\mu$m spectral range. By increasing water abundance we shift the weighting functions to lower pressures where the temperature is lower, producing a change in modelled spectrum almost indistinguishable from simply lowering the temperature. Within our optimal estimation framework we see the same high degree of anti-correlation in the retrieved covariance matrix and at first glance it might be concluded that it is impossible to distinguish between water abundance and temperature. Certainly, where optimal estimation has usually been used before it is common practice to assume the abundance of one gas to be well known such that it can be used as a `thermometer gas' and its absorption features analysed to retrieve atmospheric temperature. Since we do not, \textit{a priori}, know the water vapour abundance, how can meaningful information on temperature and abundance be retrieved? Fortunately, while the anticorrelation is strong at this pressure level for the assumed GCM-based model abundances the anticorrelation becomes less strong at higher and lower pressures due to effects such as the varying widths of the H$_2$O lines caused by pressure-broadening and the increasing importance of H$_2$--H$_2$ and H$_2$--He collision-induced absorption at depth. Hence, instead of retrieving a single value it is possible to infer a range of abundances for which we achieve similarly good fits to the observations. Analysing the secondary eclipse and primary transit observations of WASP-43b, \cite{kreidberg14} report the water vapour mole fraction to be in the range $(3.1 - 44.0)\times 10^{-4}$ from secondary eclipse observations and $(0.33 - 14)\times 10^{-4}$ from primary transit analysis, arriving at a joint constraint of $(2.4 - 21.0)\times 10^{-4}$. This estimate was further refined by \cite{stevenson17} who estimate the water vapour abundance to be  $(0.25 - 1.1)\times 10^{-4}$. The approach of optimal estimation is not best designed for exploring a wide parameter space such as this, but with care it can be used to explore the range of solutions consistent with the observations, in this particular case the synthetic observations generated from a GCM-based model atmosphere and for which we know the `ground truth'. We can see from Fig. \ref{summary} that the signatures of CO$_2$ and CO are difficult to distinguish between with the WFC3/IRAC data, as was also noted by \cite{stevenson17}. Hence, we first fixed the mole fraction of CO$_2$ to $10^{-7}$ and tried to estimate the abundance of CO only. We then set up a grid of trial CO and H$_2$O mole fractions varying between $10^{-5}$ and $5\times 10^{-3}$ and then retrieved for each grid combination the vertical profile of temperature and mean methane abundance. The \textit{a priori} methane mole fraction was set to $10^{-5}$ with the log scaling factor (natural logs) set to $0 \pm 10$  to allow the model to fit the CH$_4$ abundance as freely as possible\footnote{Remember that in optimal estimation, the \textit{a priori} errors are weights, not hard limits. Hence, it is possible to retrieve variations larger than the \textit{a priori} errors if the data are sufficiently constraining.}.
 
Figure~\ref{synthetic_ret1} shows the atmospheric parameters fitted to the synthetic phase curve data (without additional random Gaussian-generated noise) at all phases with our 1-D model, where the \textit{a priori} temperature profile was set to the nightside GCM profile with an error of $\pm 200$K. Here, we took each of the $N_{phase} = 15$ synthetic spectra in turn and for each assumed CO and H$_2$O mole fraction combination fitted the vertical temperature profile and mean CH$_4$ abundance that gave the closest disc-averaged $F_{plan}/F_{star}$ ratio (using $N_{\mu} = 5$ zenith angles and assuming azimuthal symmetry). We then took all $N_{phase}$ individually-retrieved temperature profiles and plotted them as a function of the central meridian longitude of WASP-43b (assuming 0$^{\circ}$E is the sub-stellar point). The top two rows of Fig.~\ref{synthetic_ret1} shows how $\chi^2$ varies with fixed CO and H$_2$O abundance for our models (with the second row also showing the fitted CH$_4$ abundances at $0^\circ$E), while the remaining rows plot the longitudinal variation of temperature for a single choice of the H$_2$O and CO abundance, indicated by the green asterisk in the top panel of each column.  Looking at the $\chi^2$ contour plot in Fig.~\ref{synthetic_ret1} it can be seen that we achieve an excellent fit to the synthetic observations ($\chi^2/n < 0.1$) and that the water vapour abundance is reasonably well constrained and consistent with the `true' GCM-based value. However, the CO abundance is much less well constrained. The temperature profiles shown in the first column of Fig.~\ref{synthetic_ret1} are for CO and H$_2$O values close to `true', but the remaining two columns explore how the retrieved temperatures vary with the assumed CO and H$_2$O abundances, by first increasing the assumed H$_2$O abundance by factor of 10, and then increasing the assumed CO abundance by the same factor. We find that the retrieved temperatures vary with the assumed H$_2$O abundance as expected, with increased H$_2$O abundance leading to the higher temperature regions being retrieved at lower pressure levels. However, we also find a significant variation in retrieved temperature with CO abundance, with the sub-stellar hotspot moving to higher altitude and forming a slight inversion at $\sim 0.2$ bar for large CO values; this is consistent with our observation earlier that the CO abundance and temperature at $\sim 0.1$ bar are degenerate. All the temperature profiles for the entire grid of CO and H$_2$O abundances vary moderately smoothly with longitude and altitude, indicating reasonable constraint and while they look different from the `true' temperature profiles, it would be difficult to recognise them to be in error if we did not know what the `true' values were. Hence, it can be seen that although the WFC3/IRAC data, measured with the spectral coverage and error values reported, allow some degree of constraint on the water abundance, they do not allow very strong constraint on the CO abundance due to the degeneracy with temperature.  For CH$_4$, we find that the retrieved abundances are approximately 10 times larger than `true', with retrieved error bars consistent with the spread of retrieved values. To check that the methane retrievals were not being biased by our chosen \textit{a priori} methane abundance of $10^{-5}$ we repeated our retrievals reducing this \textit{a priori} to $10^{-7}$ and found essentially identical results. The high CH$_4$ abundances retrieved with the 1-D model are consistent with one of the conclusions of  \cite{Feng16}, that the 1-D approach `overestimates both its [i.e., CH$_4$] abundance and the precision'. As for temperature, we can see  that the retrieved temperatures are hotter near the sub-stellar point, but offset eastwards as is the case in the GCM simulations used to generate the synthetic spectra. However, it can be noticed that the hotter temperatures extend some way past the day/night terminator at 90$^{\circ}$E and  90$^{\circ}$W, unlike the `true' GCM temperatures. This can be seen most clearly in the bottom left panel of Fig. \ref{synthetic_ret1}, where the retrieved--true temperatures at 0.1--1 bar show significantly negative differences of $\sim -200$~K near the sub-stellar longitude, but significantly positive differences of $\sim +200$~K at 90$^{\circ}$W and 120$^{\circ}$E. This can also be seen in Fig. \ref{GCMC}, where we compare the `true' and retrieved temperatures in detail at 180$^{\circ}$W, 90$^{\circ}$W, 0$^{\circ}$E and 90$^{\circ}$E. Here, we can see that  the 1-D model overestimates the temperature at $90^\circ$W by almost 500K. This behaviour is expected since for 1-D retrievals we use the same temperature profile at all points on the disc and to model the slow variation of observed $F_{plan}/F_{star}$ with phase angle $\zeta$, we need the temperature to vary more slowly with longitude. This can also be seen in modelled appearance of the planet in the 4.5-$\mu$m channel in Fig.~\ref{phasim}. 

Although we computed these spectra assuming our $N_\mu = 5$ Gaussian integration scheme, we checked the necessity of using such a scheme with 1-D retrievals for phase curves measured with currently achievable SNR. We did this by modelling the spectra with a single zenith angle of 45$^\circ$ and found that we achieved a similarly close fit to the observations (average differences in $\chi^2/n$, averaged over all 15 phases, of $ < 0.05$) and very similar retrieved properties. This  demonstrates that for cases where we assume the same atmospheric conditions at all observable parts of the disc, setting $\bar{R(\lambda)} = R(\lambda,\theta = 45^\circ)$ is a good approximation. In fact, we found that setting $\bar{R(\lambda)} = R(\lambda,\theta = 0^\circ)$ we were also able to achieve similarly close fits to the synthetic observations, and similar retrieved atmospheric properties as compared with the full numerical integration scheme. However, in this case, because the transmission weighting functions peak at deeper pressures the retrieved vertical temperature distribution is also shifted to deeper pressures. 

\begin{figure*}
\includegraphics[width=\textwidth]{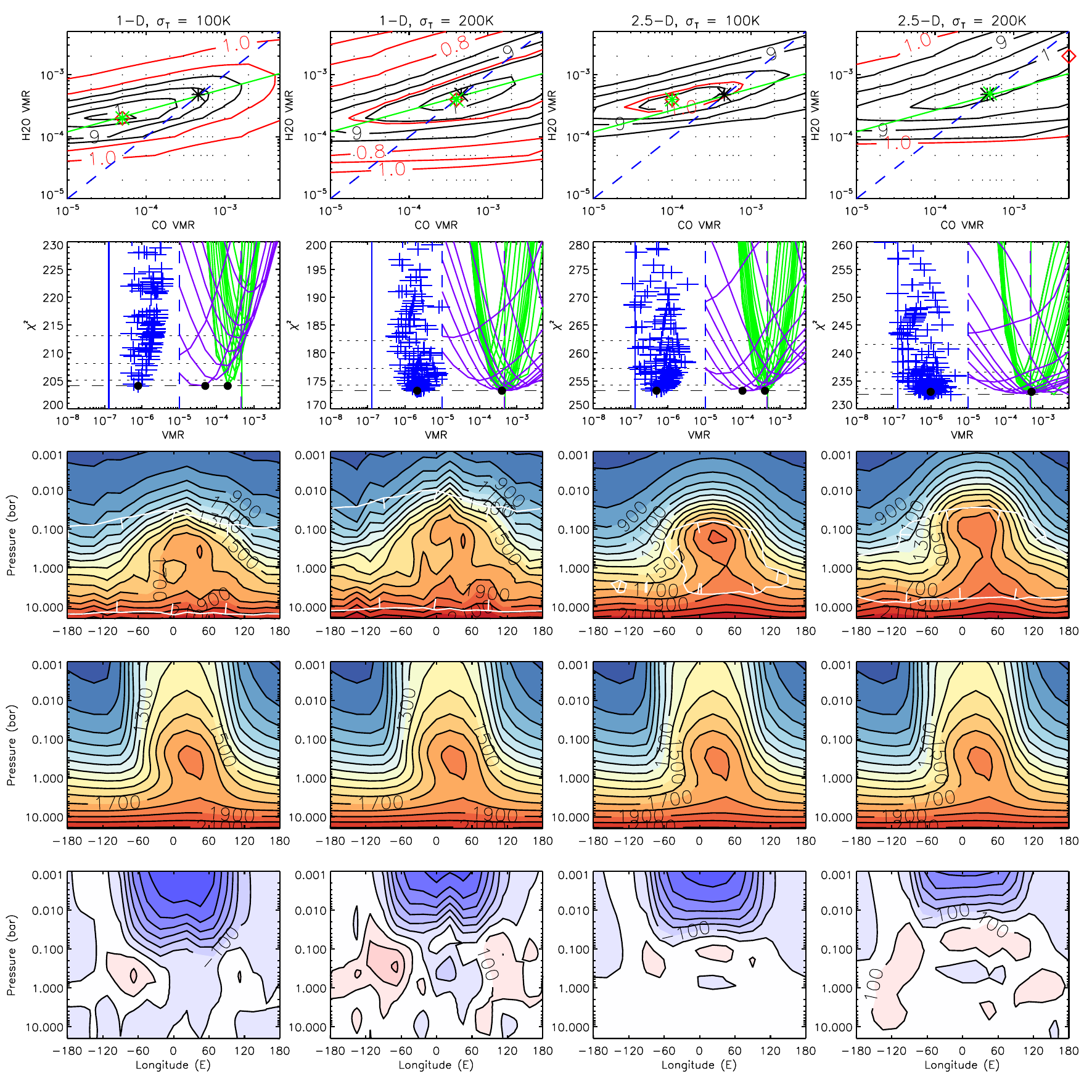}
\caption{Retrieval of GCM-based synthetic phase curve data (including randomly generated synthetic noise) using both our 1-D model and 2.5-D models. The rows are the same as described in Fig.~\ref{synthetic_ret1}. Four sets of retrievals are shown: Column 1) 1-D retrieval with \textit{a priori} temperature error $\pm 100$~K; Column 2) 1-D retrieval with \textit{a priori} temperature error $\pm 200$~K; Column 3) 2.5-D retrieval ($n=0.25$) with \textit{a priori} temperature error $\pm 100$~K; and Column 4) 2.5-D retrieval ($n=0.25$) with \textit{a priori} temperature error $\pm 200$~K. The rows are as described for Fig.~\ref{synthetic_ret1}. Here the H$_2$O and CO abundances chosen for the temperature plots are different from the ground truth and are indicted by the green asterisks in the plots on the top row. N.B., the reduced area of significant improvement for the 2.5-D retrieval with temperature error $\pm 100$~K compared with $\pm 200$~K is easily explained: by reducing the \textit{a priori} temperature error, the retrieval cannot vary as far from \textit{a priori} on the nightside since the data are less precise. Hence, the improvement factor contours surround only the dayside regions, where the data have a higher SNR. \label{synthetic_ret2}}
\end{figure*}        

While our 1-D approach gives an excellent fit to the synthetic observations and is broadly consistent with GCM-based model temperatures and abundances it is over-simplistic to assume that the atmospheric conditions are the same at all points on the disc as can clearly be seen in Fig.~\ref{phasim}. We hence applied our 2.5-D retrieval scheme (with different assumed values of the latitudinal coefficient $n$), where we fitted all $N_{phase}=15$ spectra simultaneously with a model where the state vector contained the temperature at 20 levels equally spaced in log pressure between 20 and 0.001 bar at sixteen different longitudes covering 0$^{\circ}$E to 337.5$^{\circ}$E in steps of 22.5$^{\circ}$. For the temperature profile at each longitude we again set a correlation length of 1.5 scale heights. For horizontal smoothing we applied a correlation length of 22.5$^{\circ}$ between the temperatures at different longitudes. For the CH$_4$ abundance, since we assumed horizontal `quenching' we fitted a single mole fraction for all longitudes. Figure~\ref{synthetic_ret2} compares our 2.5-D model fits to the synthetic phase curve data with our 1-D model, but this time for the synthetic phase curve data where additional random Gaussian-generated noise has been added. For our 2.5-D model we assumed a latitude temperature dependence of $(\cos\Phi)^{0.25}$ (i.e., $n=0.25$) and for both 1-D and 2.5-D retrievals we  set the \textit{a priori} temperature profile to the nightside GCM profile with errors of either $\pm 100$~K or $\pm 200$~K. It can be seen from Fig. \ref{synthetic_ret2} that we achieve similarly close fits to the synthetic phase curve data with either the 1-D or 2.5-D models, but with the 2.5-D scheme our fitted temperatures correspond much more closely with the weighted-mean longitudinal temperatures of the GCM-based model, with the hot region more tightly constrained to the dayside than the 1-D cases and so smaller differences seen the retrieved--true contour plots. To explore this more quantitatively, our fitted temperature profiles for longitudes $180^\circ$W, $90^\circ$W, 0$^\circ$E, and 90$^\circ$E are compared with the respective latitudinally-averaged GCM-based model profiles in Fig. \ref{GCMC}. Here we see that the 2.5-D-retrieved temperatures are closer to the GCM-based model profiles than the 1-D retrieved profiles and that the retrieved temperature profiles differ from \textit{a priori} by much more than the \textit{a priori} errors, underlining the point made earlier that these \textit{a priori} errors are \textbf{weights}, rather than hard limits. It can also be seen that the differences in the temperature profiles derived by the 1-D and 2.5-D retrieval models are at some pressure levels larger than the estimated retrieval error, showing these differences to be statistically significant.

With regards to composition, for the 2.5-D cases it can be seen in Fig. \ref{synthetic_ret2} that the water vapour abundance is again reasonably well constrained and consistent with the `true' GCM-based  value. However, the CO abundance is again more poorly constrained. In contrast with the 1-D cases, the 2.5-D cases recover slightly lower values of CH$_4$ with larger error bars, more consistent with the assumed GCM-based model values\footnote{In an additional test, we found that when fitting to the synthetic phase curve data without additional noise, the retrieved CH$_4$ abundances were slightly more consistent with `true'}, and again consistent with \cite{Feng16}, although the difference compared with the 1-D approach is only of 1-$\sigma$ significance at best with the assumed HST/Spitzer errors. While we have only shown 2.5-D retrievals with $n=0.25$ (i.e., $(\cos\Phi)^{0.25}$ latitude dependence) we found little difference between the temperatures and abundances retrieved by the 2.5-D scheme for latitude dependences $n=1$ and $n=2$. However, we did find that the minimum $\chi^2$ values increased monotonically as $n$ increased and that the  lowest values were found for the $(\cos\Phi)^{0.25}$ case, indicating that this is our preferred solution.  Figure~\ref{phasim} shows the fitted appearance of the disc of WASP-43b at 4.5 $\mu$m from our 1-D and 2.5-D retrievals for the case $\sigma_T = \pm 100$~K (for $n=0.25$ and $n=1$ for the 2.5-D fits) where it can be seen that the 2.5-D approach generates much more realistic-looking disc images, which correspond better with the `true' simulated appearance of this planet. It can also be seen that the $n=0.25$ case  better approximates the north/south distribution of the `true' appearance. 
We then explored how the retrieved temperatures varied with assumed CO and H$_2$O abundance for the 2.5-D approach. As for the 1-D approach we find that increased H$_2$O abundances raises the vertical location of the higher temperature regions to lower pressure levels. We also again find a significant variation in retrieved temperature with CO abundance, with the sub-stellar hotspot moving to higher altitude and forming an inversion at $\sim 0.2$ bar for large CO values. However, what is particularly clear from Fig.~\ref{synthetic_ret2} is that the best fit CO and H$_2$O values lie along a rough diagonal indicated by the green line in Figs.~\ref{synthetic_ret1} and \ref{synthetic_ret2}, and that the position along that line of the best fit depends on the assumed \textit{a priori} temperature error and thus the degree of vertical smoothing imposed by the retrieval on the vertical temperature profile. Increasing the \textit{a priori} temperature error leads not only to reduced vertical smoothing and the retrieval of temperature profiles that vary more rapidly with altitude, but also leads to the retrieval of higher H$_2$O and CO abundances. While this does not lead to very large changes in the best-fit H$_2$O abundance, it leads to large changes in the best-fit CO abundance. This suggests that trying to estimate the metallicity of WASP-43b from the CO absorption for the WFC3/IRAC data is likely to be prone to significant systematic error since the inferred value depends greatly on the assumed/fitted temperature profile.  In simple terms this correlation arises from the fact that, as we can see from Fig. \ref{summary}, the most sensitive wavelength for CO is the 4.5-$\mu$m Spitzer channel, while the temperature solution is driven by fitting to all the channels simultaneously. Hence, the optimal estimation approach finds that that when fitting to the observations with a temperature profile that is constrained to not to vary too greatly with altitude, a lower CO abundance is required in order to allow the 4.5-$\mu$m channel to see to the deeper, warmer levels, where the radiances are higher. If the temperature profile is able to vary more rapidly, higher temperatures are retrieved at the lower pressures and hence higher CO values are needed to push the weighting function for the  4.5-$\mu$m channel up to pressure levels where the radiances are consistent with the observations. A similar effect is seen for the CH$_4$ retrievals, for which the most sensitive wavelength is the 3.6-$\mu$m Spitzer/IRAC channel.  It should be noted that this correlation between CO and H$_2$O abundance has been seen in previous retrieval analyses using Bayesian approaches \citep[e.g.,][]{kreidberg14,line16}.

For both 1-D and 2.5-D retrievals, we see that the difference between retrieved and `true' is worst for pressures less than 0.05 bar on the dayside. This is because the retrievals tend towards the \textit{a priori} estimates at pressures where there is little information. Figures \ref{synthetic_ret1} and \ref{synthetic_ret2} also plot  contours of the temperature `Improvement factor', $I$, for the temperature retrievals. This is defined as $I = 1-\sigma_r/\sigma_a$, where $\sigma_a$ is the \textit{a priori} error and $\sigma_r$ is the retrieval error. Higher values of $I$ indicate where the retrieval is more constrained by the measurements and thus where the retrieved temperatures are most reliable. We can see here that this region corresponds to the $\sim 0.05 - 10$ bar region for the 1-D approach and  $\sim 0.2 - 10$ bar region for the 2.5-D approach (with diminished improvement factors on the nightside for the 2.5-D retrievals since the temperatures are lower here leading to reduced SNR and solutions that can deviate less from the \textit{a priori} estimate). Outside this pressure region, the temperature profiles revert smoothly to \textit{a priori} and thus differences between `true' and retrieved should not cause concern. We repeated our retrievals using our other \textit{a priori} temperature profile ($T=1000$~K at all altitudes) and found consistent results within the pressure regions where the `Improvement factor' was significant. 

\begin{figure*}
\includegraphics[width=\textwidth]{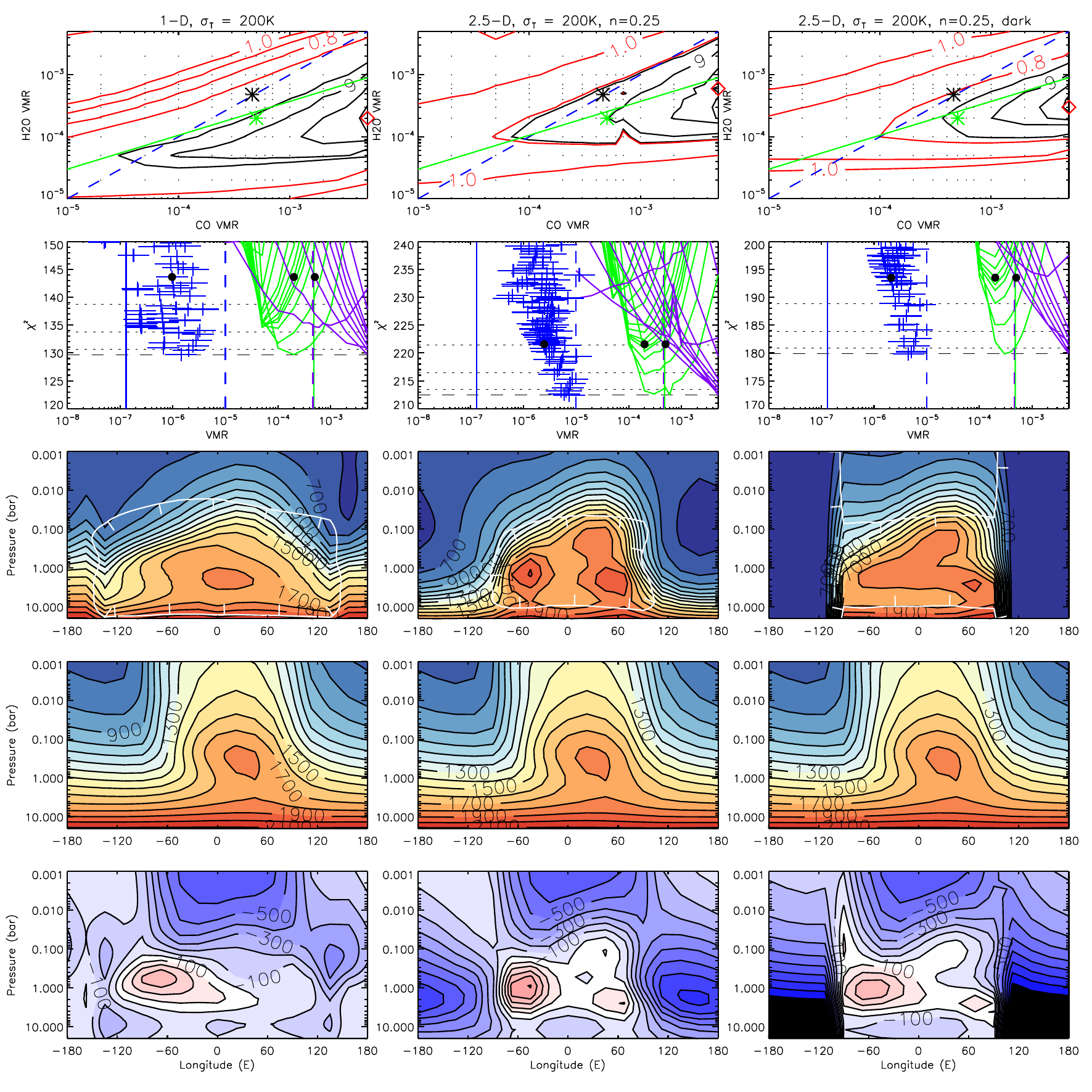}
\caption{Application of our 1-D and 2.5-D models ($n$=0.25) to the measured phase curves of \citet{stevenson17}, assuming an \textit{a priori} temperature error of $\pm 200$~K. The rows are as described for Fig.~\ref{synthetic_ret1}. For the temperature maps, we have chosen H$_2$O and CO abundances (indicated by green asterisks in row 1) that balance the requirement to have the lowest $\chi^2$ values with the expectation to have roughly equal abundances ($2 \times 10^{-4}$ and $5 \times 10^{-4}$, respectively for all cases). \label{meas_ret1}}
\end{figure*}

\begin{figure*}
\includegraphics[width=\textwidth]{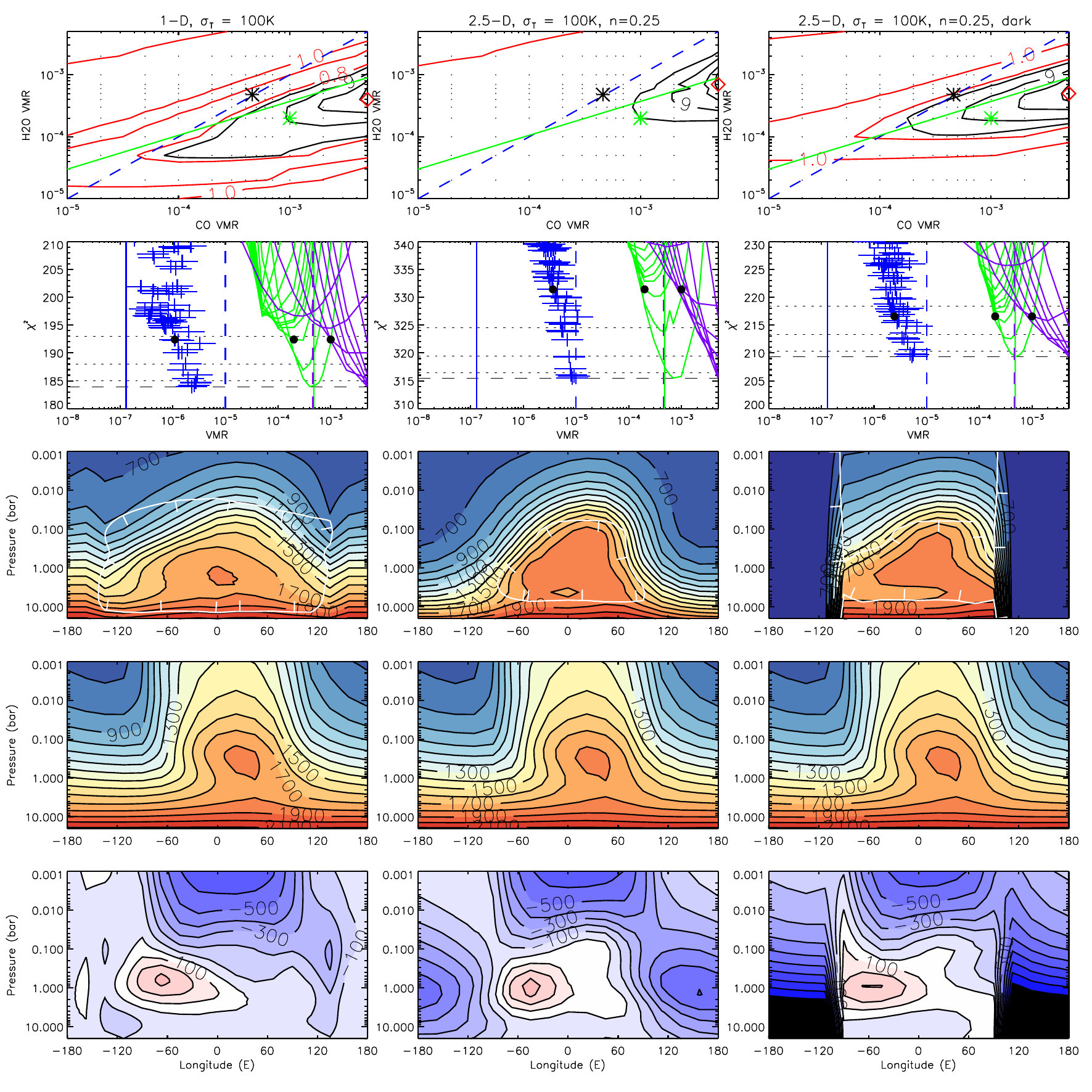}
\caption{Application of our 1-D and 2.5-D models ($n$=0.25) to the measured phase curves of \citet{stevenson17}, assuming an \textit{a priori} temperature error of $\pm 100$~K. The rows are as described for Fig.~\ref{synthetic_ret1}. For the temperature maps, we have chosen H$_2$O and CO abundances (indicated by green asterisks in row 1) that balance the requirement to have the lowest $\chi^2$ values with the expectation to have roughly equal abundances ($2 \times 10^{-4}$ and $1 \times 10^{-3}$, respectively for all cases). \label{meas_ret2}}
\end{figure*}

\begin{figure*}
\includegraphics[width=\textwidth]{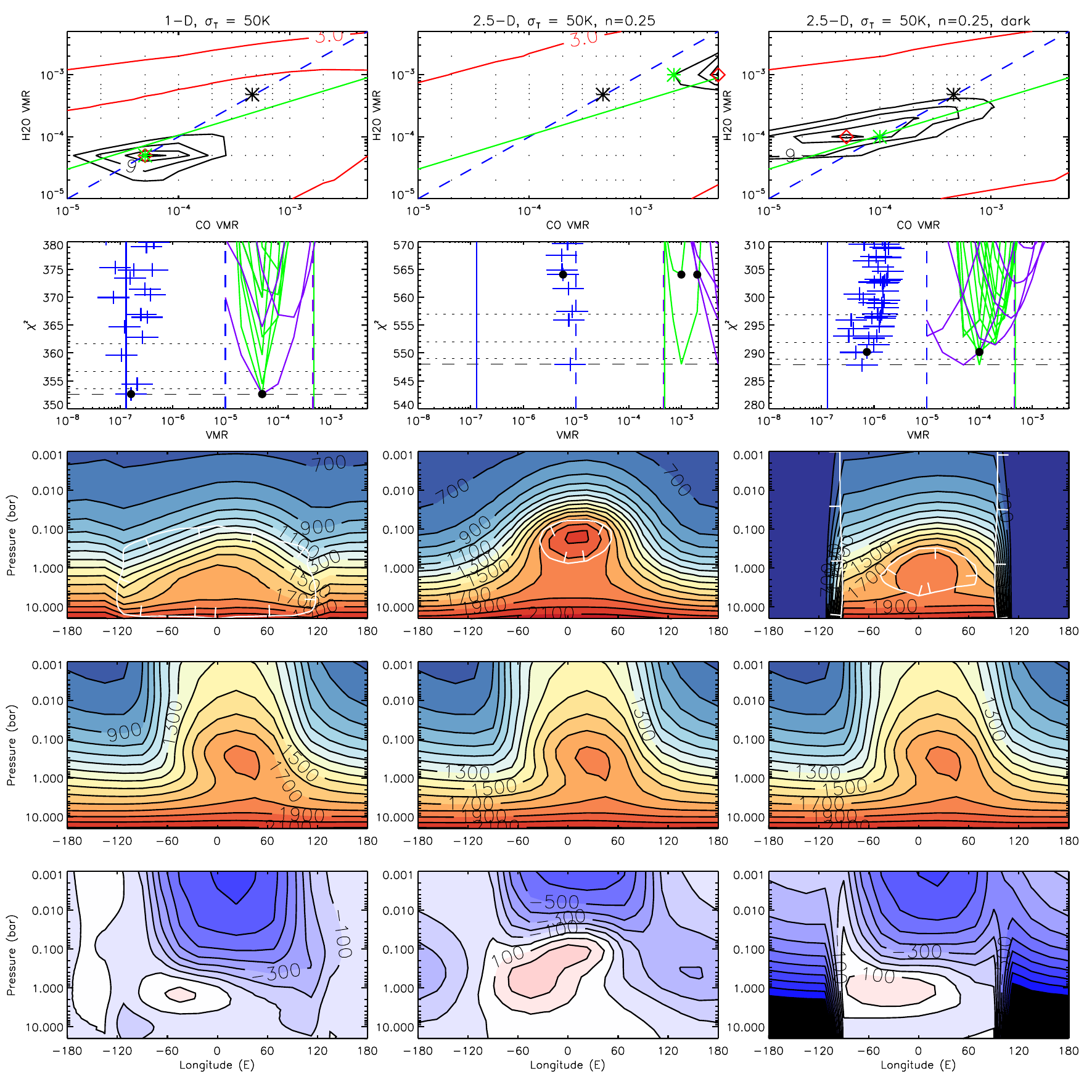}
\caption{Application of our 1-D and 2.5-D models ($n$=0.25) to the measured phase curves of \citet{stevenson17}, assuming an \textit{a priori} temperature error of $\pm 50$~K. The rows are as described for Fig.~\ref{synthetic_ret1}. For the temperature maps, we have chosen H$_2$O and CO abundances (indicated by green asterisks in row 1) that balance the requirement to have the lowest $\chi^2$ values with the expectation to have roughly equal abundances. In this case, rather different solutions are chosen for the three cases shown. \label{meas_ret3}}
\end{figure*}

\section{Application of retrieval schemes to real WFC3+IRAC observations}

Having demonstrated the efficacy and reliability of our modelling approach on a  set of simulated observed data, where we knew what the `ground truth' was, we then applied both our 1-D and 2.5-D models (using identical setups to the validation tests) to the observed phase curves of \citet{stevenson17}. Our results are shown in Figs.~\ref{meas_ret1}, \ref{meas_ret2} and \ref{meas_ret3} for \textit{a priori} temperature errors of $\pm 200$~K, $\pm 100$~K and $\pm 50$~K, respectively.

Comparing the 1-D and 2.5-D retrievals it can be seen that the nightside of WASP-43b appears much colder than the GCM-based simulation and that the 2.5-D approach finds that the hot region of the planet is much more tightly constrained to the dayside of the planet than the 1-D approach. For the dayside we retrieve the highest temperatures at $\sim 200$ mb, significantly shifted to the east, but at lower altitudes the hotter regions appear to be centred more to the west, mirroring the GCM simulations, where the hotter regions are shifted more to the east near the equator, but further westward at mid to polar latitudes (Fig.~\ref{GCMA}). It seems in our retrievals that this pattern may be even more enhanced in the actual atmosphere of WASP-43b. This day-night temperature difference becomes much more marked for the 2.5-D case where the modelling finds the temperature at all longitudes simultaneously. We repeated these retrievals for different values of the latitude dependence coefficient, $n$, and as for the GCM simulations found the best fits for $n=0.25$. Figure~\ref{phasim} shows the fitted appearance of the disc of WASP-43b at 4.5 $\mu$m from our 1-D and 2.5-D retrievals for the case $\sigma_T = \pm 100$~K (for $n=0.25$ and $n=1$) where it can be seen that the 2.5-D approach generates much more realistic-looking disc images, which correspond much better with the expected appearance of this planet. 

\begin{figure}
\includegraphics[width=\columnwidth]{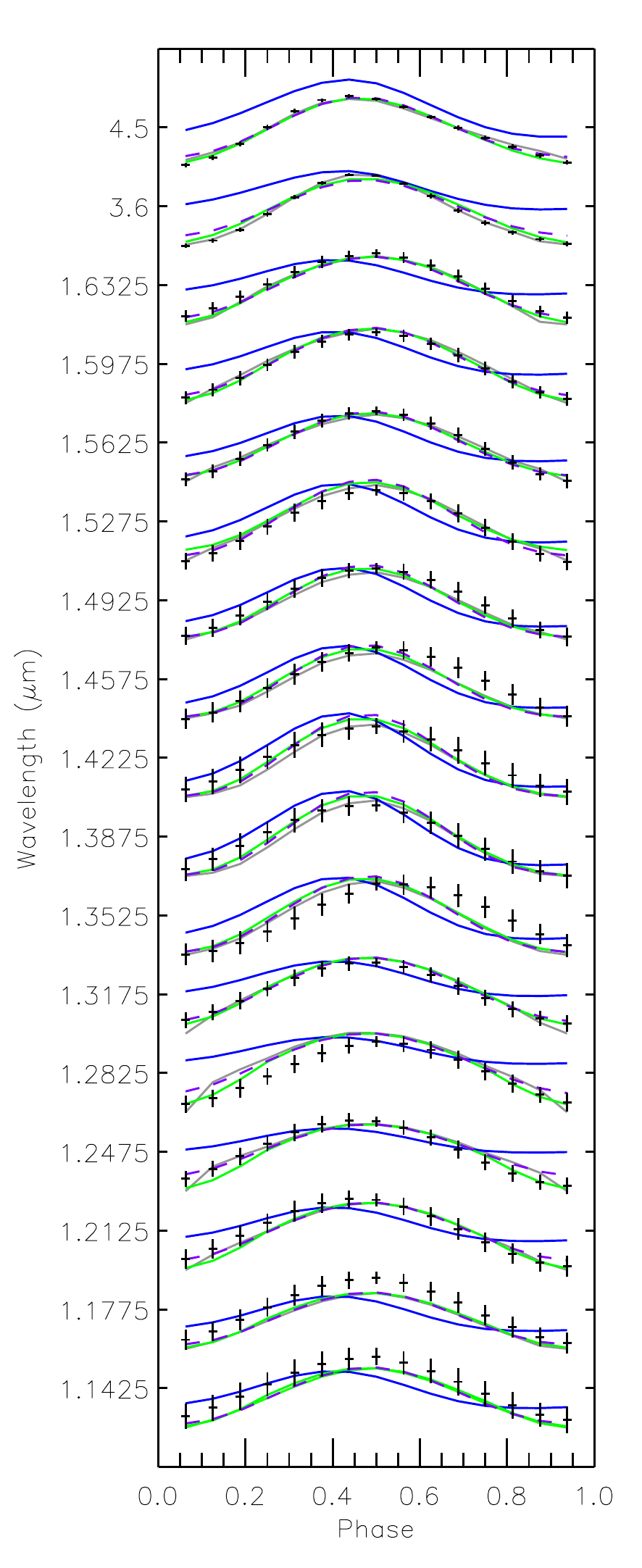}
\caption{Fit to the phase curves ($\sigma_T = 100$~K) for each wavelength bin reported by \citet{stevenson17}, normalised for ease of reference. The measured, normalised phase curves and errors are indicated by the points and error bars. The line colours are: grey -- fits to these curves using our 1-D model; purple-dashed -- fits to these curves using our 2.5-D model, assuming $(\cos \Phi )^{0.25}$ latitude dependence; and green -- fits to these curves using our 2.5-D model, assuming $(\cos \Phi )^{0.25}$ latitude dependence and forcing nightside temperatures to 500~K. For reference, the normalised synthetic phase curves caclulated from our GCM-based model atmosphere are shown in blue. \label{comparephase}}
\end{figure}

\begin{figure*}
\includegraphics[width=\textwidth]{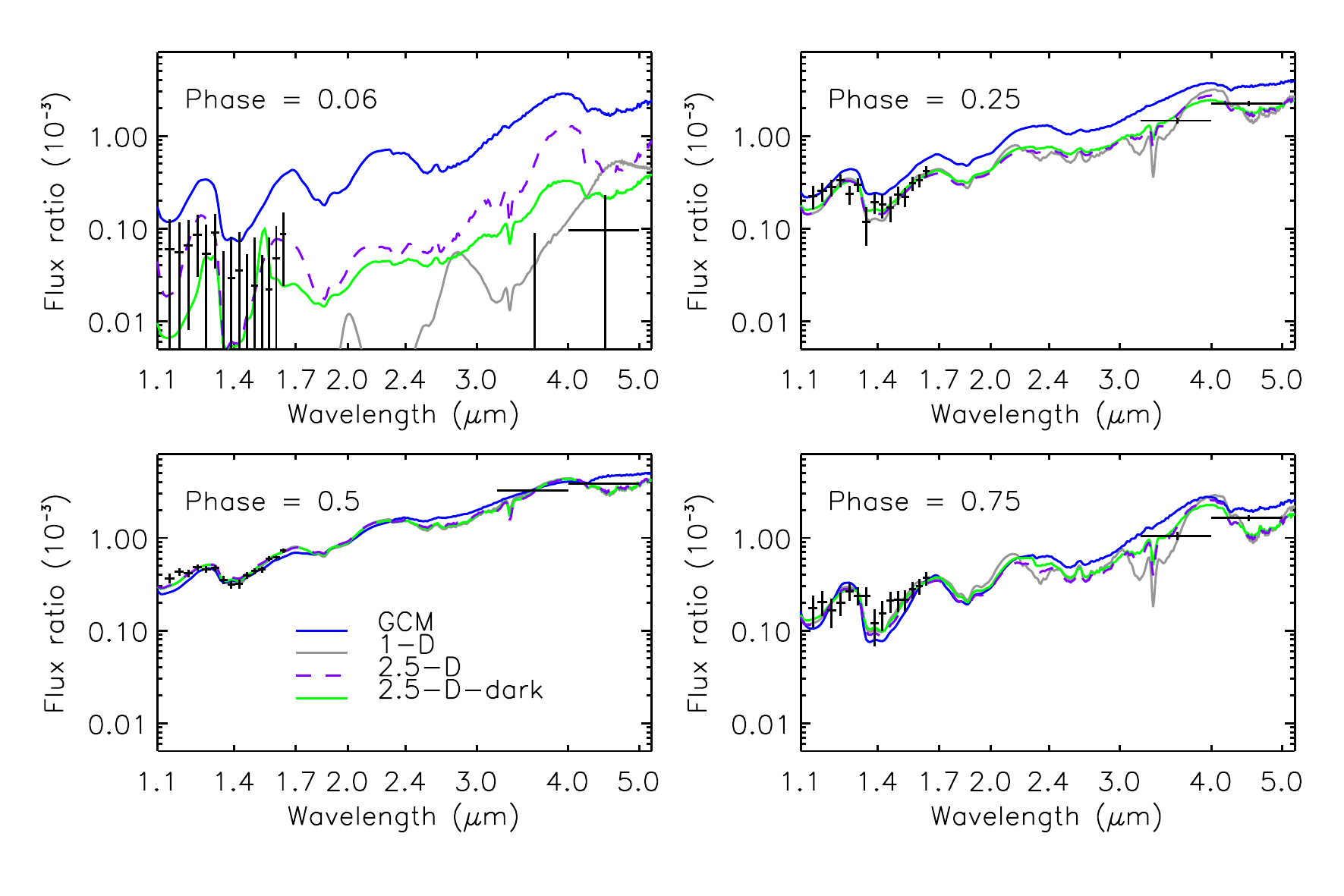}
\caption{Phase curves for each wavelength bin reported by \citet{stevenson17}, compared with moderate resolution spectra ($\Delta\lambda = 0.02$ $\mu$m) calculated from the fitted atmospheric temperatures and abundances ($\sigma_T = 100$~K) for the same geometry. Curves are compared for phases 0.06, 0.25, 0.5 and 0.75. The observations are compared to the \textit{ab initio} GCM-based simulations (blue), our 1-D model fit (grey), our 2.5-D model, assuming $(\cos \Phi )^{0.25}$ latitude dependence (purple - dashed) and our 2.5-D model, assuming $(\cos \Phi )^{0.25}$ latitude dependence and forcing nightside temperatures to 500~K (green).  \label{comparephase_qt}}
\end{figure*}

The much colder temperatures retrieved on the nightside of WASP-43b (especially for the 2.5-D case), which is difficult to reproduce in GCM simulations, means that emission from the nightside contributes much less to the disc-averaged radiance for phases that can see both sides of the planet. Although the Improvement Factors are found to be lower on the nightside than the dayside for the 2.5-D retrieval, we should not think that the retrieval does not return information on the nightside. Rather its best solution comes from driving the nightside temperatures so low that it cannot specify exactly how cold they should be, only that they should be low enough not  to contaminate too much the radiance from the dayside. This phenomenon has been noted before \citep{kataria15, parmentier16} and it has been used to suggest that if the temperatures fall low enough on the nightside that a thick cloud forms at a level which effectively obscures any emission. From the transmission weighting functions shown in Fig.~\ref{summary} we can see that such a cloud must have a cloud top pressure less than $\sim 200$ mb in order to suppress emission at all the WFC3/IRAC wavelengths. To explore whether such a scenario could be supported within our 2.5-D retrieval framework, we repeated our retrievals with the same temperature \textit{a priori} on the dayside, but fixing the temperatures to be arbitrarily low (500~K) on the nightside, so that we only retrieved the dayside temperature variation and dayside abundances of H$_2$O, CO and CH$_4$. Our results are shown in column 3 of Figs.~\ref{meas_ret1} -- \ref{meas_ret3} for $\sigma_T = \pm 200$~K, $\pm 100$~K and $\pm 50$~K, respectively. Here, it can be seen from the fitted $\chi^2/n$ values that we achieve a much closer fit to the data when we force the nightside emission to be very low than when we try to fit the temperature on both dayside and nightside. This suggests that in the former retrievals the temperatures on the nightside were limited from falling as low as the retrieval model wanted by the \textit{a priori} constraints. By forcing the nightside to be very cold (and thus simulating the presence of a cloud with a top at $< 200$ mb and cloud-top temperature of $\sim 500$K) we obtain an excellent fit to the observations and can also place tighter constraints on the gaseous abundances with the H$_2$O mole fraction found to be in the range $(2 - 10) \times 10^{-4}$. Hence, we conclude that the WASP-43b phase curve observations are consistent with the nightside of this planet being obscured by cloud. The simulated appearance of WASP-43b at 4.5 $\mu$m with this `cloudy' 2.5-D model is also shown in Fig.~\ref{phasim}. Figures~\ref{comparephase} and \ref{comparephase_qt} show that our retrievals (for $\sigma_T = 100$~K) achieve very close fits to the observed phase curves of WASP-43b. 

Although the retrieved temperatures appear plausible and the observations are consistent with the presence of cloud on the nightside, the gas abundances inferred from these observations is more puzzling. It can be seen that for all retrievals from the real observations of WASP-43b \citep{stevenson17} a CO abundance of $5\times 10^{-3}$ is favoured, unless the \textit{a priori} temperature error is reduced to $ < 100$~K,  in which case the stronger \textit{a priori} constraint leads to noticeably poorer fits and higher $\chi^2/n$ values. We also find that the methane abundance lies in the range $(1-10) \times 10^{-6}$ for the 2.5-D cases. At high temperatures (such as those found in the atmosphere of WASP-43b) we expect carbon to be predominantly in the form CO and oxygen to be split between CO and H$_2$O. We thus expect the CH$_4$ abundance to be very low at such temperatures. For a solar composition atmosphere the C/O ratio is  $\sim 0.53$  and so we expect roughly equal abundances of H$_2$O and CO and very low CH$_4$ abundance, which is what is found in the GCM-based model. However, for the WASP-43b observations we find the best fit CO and CH$_4$ abundances to be ~10 times greater than expected, from which we infer a C/O ratio of $\sim 0.91$, since CO/H$_2$O $\sim$ C/(O-C). The question of the C/O ratio in the atmospheres of giant exoplanets has been addressed by a number of authors \cite[e.g.,][]{oberg11,madhusudhan14,espinoza17,cridland19}. When planetesimals form they trap more oxygen than carbon in solid form, meaning that the gas is enriched in C/O. Authors such as \citet{oberg11} argue that this means that giant planets should be carbon-rich since they accrete more gas than planetisimals, but authors such as \citet{espinoza17} note that if such planets accrete both gas and solids at the same time, then they should generally be oxygen-rich. Hence, there is considerable interest in the C/O ratio of planetary atmospheres and so our retrieved C/O ratio is potentially significant. However, the derivation of this ratio is prone to several sources of systematic error, which need to be fully addressed. One possible source of systematic error we considered was whether we might be using an incorrect stellar spectrum (needed to convert the measured $F_{plan}/F_{star}$ observations to radiance). However, comparing our stellar spectrum from the Kurucz ATLAS model atmospheres \citep{castelli2004new} with the stellar spectrum for WASP-43b assumed by other authors from PHOENIX \citep{allard00} we found insignificant differences. Instead, we believe our main source of error in determining C/O is that the abundance of CO comes almost entirely from the 4.5-$\mu$m Spitzer/IRAC observations and the CH$_4$ abundance from the 3.6-$\mu$m Spitzer/IRAC observations, while the temperature comes from the combined HST/Spitzer observations. We already noted earlier that there is considerable degeneracy between atmospheric temperature near the tropopause and the abundances of these two constituents. However, in addition to this, we must remember that if there is any systematic error in the $F_{plan}/F_{star}$ ratios between the HST/WFC3 and Spitzer/IRAC observations, we could very easily retrieve systematically offset CO abundances. For example, if the Spitzer radiances flux ratios are either lower than they should be compared with the HST observations or, alternatively, if the HST flux ratios are higher than they should be, then the optimal fit to the combined data would be obtained by increasing the CO and CH$_4$ abundance to lower the Spitzer radiances relative to the HST radiances, which may be what is happening here. We note that in two independent reprocessings of the Spitzer flux ratios \citet{mendonca18} and \citet{morello19} found significantly different values from \citet{stevenson17} for the Spitzer/IRAC channels, which suggests that there may indeed by systematic differences between the HST/WFC3 and Spitzer/IRAC channels in the reported phase curve spectra. The fact that we find both CO \textbf{and} CH$_4$ to be larger than we might expect for a hot solar composition atmosphere leads us to suspect that these differences do exist. Finally, we note that the larger CO abundances we infer here compared with the lower abundances retrieved from the same data by \citet{stevenson17} are most likely caused by the differences in temperature profile parameterisation. In this work we have allowed the atmospheric temperature to vary at all altitudes and we have seen that there is a degeneracy between how rapidly the temperature is allowed to vary with altitude and the retrieved CO abundance. The temperature profile used in retrieval model employed by \citet{stevenson17}, CHIMERA \citep{line14}, is parameterised (as is necessary for Bayesian models that explore a large parameter space) and so has an implicit limitation in vertical variability; any such constraint on vertical variability could lead to systematic differences in the inferred CO abundance. 

Although the best fits are for rather large CO values, it can be seen from Figs.~\ref{meas_ret1} -- \ref{meas_ret3} that a wide range of CO and H$_2$O values gives fits to the data with a $\chi^2/n < 1.0$. If we assume that the metallicity of WASP-43b is similar to that of the Sun then we would expect the abundances of H$_2$O and CO to be very similar and thus the solution to lie along the dashed blue lines in the $\chi^2$ contour plots of Figs.~\ref{meas_ret1} -- \ref{meas_ret3}. Looking at the right-hand column (i.e., where the nightside emission has been forced to be very low) for Fig.~\ref{meas_ret2}, for $\sigma_T = 100$~K, it can be seen that solutions where the mole fraction of both H$_2$O and CO are in the range $(1 - 7) \times 10^{-4}$ fit the observed spectra to $\chi^2/n < 0.9$. For these values, the mole fraction of CH$_4$ is $(1 - 3) \times 10^{-6}$. Since we fix  the abundance of CO$_2$ to $10^{-7}$ our estimate for CO+CO$_2$ is effectively the same as that for CO.

For comparison, with their 1-D Bayesian approach and parameterised temperature profiles, \citet{stevenson17} report mole fractions of $10^{(-4.28 \pm 0.32)} = (0.25 - 1.1)\times 10^{-4}$ for H$_2$O, $10^{(-3.5 \pm 0.4)} = (1.3-7.9) \times 10^{-4}$ for CO+CO$_2$, and  $>  10^{-5.3} = 5\times 10^{-6}$ for CH$_4$. These results overlap with our determinations, but we find higher abundances of H$_2$O are consistent with the observations. 

\section{Discussion}

We have developed a novel retrieval technique for modelling the phase curves of exoplanets that we have shown is robust and which provides an excellent fit to the observations of WASP-43b reported by \citet{stevenson17}.   Our approach uniquely models the radiance contributed to the disc-averaged radiance from both dayside and nightside simultaneously with a full treatment of observation angles and disc integration. We find that the nightside temperatures are required to be much cooler than those predicted from the traditional 1-D approach to be consistent with observations and our fitted temperatures and gaseous compositions agree much better with the `true' values of our GCM-based model atmosphere when tested on GCM-based synthetic observations.  Indeed, we find nightside temperatures of WASP-43b to be so low that we can model the observations effectively from the dayside radiances only, suggesting that the nightside of WASP-43b may be obscured by a thick cloud with a cloud top at $< 0.2$ bar.

We also find considerable degeneracy in our retrievals between gaseous abundance and temperature. While the abundance of water vapour is reasonably well constrained, the retrieval of the abundance of CO and CH$_4$ comes effectively from the two Spitzer/IRAC observations at 3.6 and 4.5 $\mu$m, co-retrieved with temperatures derived to be most consistent between the HST and Spitzer channels. It is found that allowing different degrees of vertical smoothing in the retrieved temperature profiles gives substantial differences in the retrieved CO abundance. This behaviour helps to explain why our retrievals appear to favour considerably more CO than the study of \cite{stevenson17}, which used a temperature profile that was more vertically constrained. We do not believe the difference is caused by systematic radiative transfer differences in the NEMESIS model used here and the CHIMERA model \citep{line14} used by \cite{stevenson17}, since a recent study \citep{barstow18} has shown that these models generate consistent spectra given identical inputs. However, we note that possible systematic differences between the radiometric calibration between the HST/WFC3 and Spitzer/IRAC datasets could easily lead to large differences in the retrieved CO/H$_2$O ratio.

Although our approach is a significantly more geometrically realistic than previous methods, there are potentially a number or areas in which it could be improved and/or developed, which we will discuss now.

1) We used an optimal estimation retrieval model here because such an approach is fast, especially in implicit differentiation mode (where the Jacobians are calculated analytically rather than by numerical differentiation), which NEMESIS can employ for thermal emission calculations. In addition, we used channel-averaged k-tables, which gives an additional improvement in computational speed. Even then, however, the retrievals are time consuming as we have to model $N_{phase} \times N_{wave} = 15 \times 17 = 255$ flux ratio measurements using a disc-integration scheme that for $N_\mu = 5$ comprises 45 -- 47 points on the disc for each phase angle. Hence, $255 \times 45 = 11,475$ radiative transfer calculations are done for each iteration of the model. Since our optimal estimation model typically takes 10 -- 15 iterations to converge, this gives $\sim$ 138,800 calculations per retrieval. While the optimal estimation method is fast and efficient, it is less easy to explore the full solution space than Bayesian techniques such as Monte-Carlo Markov Chain or Nested Sampling. However, in this study we have mapped a wide range of acceptable solutions by performing retrievals for a grid of assumed H$_2$O and CO abundances and find that such an approach allows us to give conservative estimates on the likely gaseous abundances and leads to more realistic-looking temperature distributions than the 1-D approach. Nevertheless, although Bayesian techniques need $\sim$ 10,000 iterations before they converge, which is considerably more than 10 -- 15 iterations needed for optimal estimation, we plan in the near future to incorporate the new 2.5-D approach into our Nested Sampling version of NEMESIS to more fully explore solutions that are consistent with the observed phase curves.

2) We have modelled these observations assuming no clouds. It is possible that the reason we see such a large day/night contrast in the WASP-43b observations is not that the temperatures vary greatly with longitude, but that the nightside, which is cooler, forms clouds that obscure the thermal emission from lower altitudes \citep[e.g.,][]{kataria15, parmentier16}. We have assessed this possibility with our model and find that it gives improved fits to the observations. As we note, looking at the weighting functions in Fig.~\ref{summary} it can be seen that such a cloud would have a cloud top at pressure level of $<0.2$ bar in order to suppress the radiation at all 17 wavelengths considered here, with a cloud top temperature of $\sim 500$K. 
In this paper we have modelled the effect of clouds by forcing the nightside temperatures to arbitrarily low values. However, in future work we could instead add a cloud into our radiative transfer scheme. To do this, we would need to make some estimate of its complex refractive index spectrum in order to compute, with Mie theory, its extinction cross-section spectrum. Since we need a cloud that is opaque over the whole range considered here of 1 -- 5 $\mu$m, it is likely that the particles must have radii on the order of a few $\mu$m.

3) If we are in the future to include clouds in our model, we should then also consider the scattered stellar contribution they might make. We noted earlier and showed in Fig.~\ref{summary} that reflection from cloud could potentially make a significant contribution to the observed flux ratios at wavelengths less than 2 $\mu$m. We have designed our disc-averaging scheme to be able to incorporate scattering, but such computations are hugely computationally challenging. Not only are the radiative transfer calculations themselves computationally expensive, but our implicit differentiation scheme is inefficient for tracking the partial derivatives of our Matrix-Operator scattering model \citep{plass73} and hence the functional derivatives are calculated numerically. In our current model we have 16 longitudes, 20 pressure levels for temperature and 4 gas abundances giving 384 state vector elements. Hence, if calculating the functional derivatives numerically, our radiative transfer calculations would be at least 385  times slower, even before factoring in the at least 10 times slower speed of our scattering model. Hence, such modelling will be challenging.

4) We have here assumed the temperature profile to be continuous and retrieve at all pressures in our model. The use of a parameterised profile would simplify and further increase the speed of our model, but at the disadvantage of restricting the range of solutions we can explore to those covered by the parameterisation. Such an approach could be optimal for Bayesian realisations of our 2.5-D retrieval technique, but as we have seen there is considerable degeneracy between temperature and the abundance of CO and H$_2$O. By constraining the temperature profile more strongly, we may bias our estimates of the abundances of the gaseous constituents of the atmosphere of WASP-43b. It can also be seen from Figs. \ref{meas_ret1} - \ref{meas_ret3} and previous Bayesian retrivals from these data \citep[e.g.,][]{stevenson17}, that the prior and method chosen to fit to the observations can significantly change the retrieved temperatures also. Hence, the accuracy with which we can estimate how the temperature varies with longitude and altitude and also the molecular abundances is strongly limited by the inherent degeneracy in these HST/WFC3 data. 

5) We have assumed here that there is a single $n$ coefficient the $(\cos\Phi)^n$ latitude dependence. However, looking at the brightness of the disc simulated from the GCM atmosphere in Fig.~\ref{phasim} it can be seen that while $(\cos\Phi)^{0.25}$ is generally most appropriate, at some phases $\cos\Phi$ or $(\cos\Phi)^2$ might be a better approximation. It would be interesting to explore whether a parameterisation that combines different $n$-dependencies might give more reliable retrievals. 

6) Finally, in this study we assumed the abundance of CH$_4$ to be the same at all longitudes for the 2.5-D model, because of the expected horizontal `quenching'. However, our model setup is also able to retrieve abundances that do vary with longitude, as noted earlier. In early tests with such a model using synthetic phase curves generated from GCM simulations that included longitudinal abundance variation (rather than the hybrid test model used here, where the abundances were set to longitudinal averages) we found that we were able to recover significant longitudinal variation in CH$_4$, approximately matching the synthetic model variation. However, adding this additional degree of freedom led to some instability in the temperature retrievals, which is easily understood given the degeneracy between CH$_4$ abundance and temperature at pressures greater than 0.1 bar, already noted. However, future measurements covering a wider wavelength range at higher SNR might be able to break such degeneracy, and we could then use this approach to determine longitudinal variations in abundance also.

\section{Conclusion}

We have developed a novel retrieval technique, which we call a 2.5-D retrieval approach that attempts to model more realistically and reliably phase curves observations of exoplanets. In our 2.5-D approach we retrieve the vertical temperature profile and mean gaseous abundance at all longitudes and latitudes \textbf{simultaneously}, assuming that the temperature or composition, $x$, at a particular longitude and latitude $(\Lambda,\Phi)$  is given by $x(\Lambda,\Phi) = \bar{x} + (x(\Lambda,0) - \bar{x})\cos^n\Phi$, where $\bar{x}$ is the mean of the morning and evening terminator values of $x(\Lambda,0)$, i.e., $\bar{x} = (x(-90^\circ,0) + x(90^\circ,0))/2$, and $n$ is an assumed coefficient. We find that this model fits the synthetic flux ratio phase curve observations of our WASP-43b GCM-based simulations more realistically than the traditional 1-D approach, and provides a more geometrically realistic retrieval of the observed WASP-43b phase curve observations of  \citet{stevenson17}. Our retrieved gaseous abundances are broadly consistent with previous determinations and we find the mole fraction of H$_2$O lies in the range $(1-10)\times 10^{-4}$, which is consistent with the atmosphere of this planet having an approximately solar composition. In tests where we retrieve from synthetic phase curves generated from a GCM-based model, we find that the current HST/Spitzer WASP-43b data provide much poorer constraint on the abundance of CO, with the choice of \textit{a priori} temperature profile influencing the best fit parameters. Finally, when fitting to the measured phase curves reported by \citet{stevenson17} we find that the best-fit CO and CH$_4$ abundances both seem biased to larger values than we might expect, with CH$_4$ $\sim 10$ times larger than the expected thermodynamically-expected value and CO/H$_2$O $\sim 10$, suggesting a C/O ratio of $\sim 0.91$. While the CO/H$_2$O results alone might lead us to conclude than C/O $\sim$ 1, the fact that the CH$_4$ abundance also seems overestimated leads us to suspect that there might be a systematic difference in the radiometric calibrations between the HST/WFC3 and Spitzer/IRAC observations. We conclude that the uncertainty on the Spitzer points is probably underestimated by the error bars assumed in these retrievals as three different data reductions of the same phase curve \citep{stevenson17,mendonca18, morello19} lead to flux ratio differences larger than 1-$\sigma$. Since the inferred C/O ratio is highly dependent on the relative HST versus Spitzer flux ratios (and the errors on these ratios) a robust determination will not be possible until a consensus is arrived at in the community on the radiometric calibrations of these data sets, which includes all possible contaminating effects \citep[e.g., stellar variability,][]{knutson12}. 

Having demonstrated the efficacy of this 2.5-D approach for the HST/Spitzer phase curve observations of WASP-43b, the next step will be to apply them also to recent observations of WASP-103b \citep{kreidberg18} and WASP-18b \citep[e.g.,][]{arcangeli19} and possible new observations of other exoplanets. Further into the future, the James Webb Space Telescope (JWST) plans to make phase curve observations of several exoplanets over a wide wavelength range with much higher SNR than the HST/Spitzer observations analysed here. It is hoped that these improved observations will break many of the degeneracies inherent in existing observations and allow us to retrieve temperature profiles that are much less dependent on the prior and method chosen to fit to the observations. In particular, the acquisition of high SNR observations covering the absorption features of H$_2$O, CO and CH$_4$ should allow us to more effectively discriminate between the abundances of these gases and temperature in the atmosphere of WASP-43b and other exoplanets and thus enable us to more reliably determine the C/O ratios and enrichment of heavy elements in these exotic worlds. The new 2.5-D approach described here should be particularly effective in analysing these improved data and we intend to start testing this approach on synthetic JWST observations in the near future.

\section*{Acknowledgements}

This work was supported by United Kingdom's Science and Technology Facilities Council. The authors would like to sincerely thank the anonymous reviewer for their careful and detailed review, and very helpful suggestions on how to improve the presentation of this paper.




\bibliographystyle{mnras}
\bibliography{bibliography} 








\bsp	
\label{lastpage}
\end{document}